\documentclass[reprint,superscriptaddress,pra,nofootinbib,preprintnumbers,floatfix]{revtex4-1}
\usepackage{latexsym,amsfonts,bm}
\usepackage{xspace}
\usepackage{cancel}
\usepackage{graphicx}    
\usepackage{microtype}   
\usepackage{amsmath}     
\usepackage{amsthm}    
\usepackage{amssymb}   
\usepackage{siunitx}     
\usepackage{booktabs}    
\usepackage[colorlinks=true,citecolor=blue]{hyperref}
\usepackage[capitalize]{cleveref}
\usepackage{tensor}

\usepackage{adjustbox}    
\usepackage{subcaption}
\usepackage{xcolor}
\usepackage{mathtools} 
\usepackage{bm}
\usepackage{dsfont}
\usepackage[shortlabels]{enumitem}  
\usepackage{physics}   

\providecommand{\ie}{\emph{i.e.}}
\providecommand{\eg}{\emph{e.g.}}

\providecommand{\CE}{\mathcal{E}}

\providecommand{\CR}{\mathcal{R}}

\providecommand{\uone}{$\mathrm{U}(1)$\xspace}
\providecommand{\sutwo}{$\mathrm{SU}(2)$\xspace}
\providecommand{\suthree}{$\mathrm{SU}(3)$\xspace}
\providecommand{\sothree}{$\mathrm{SO}(3)$\xspace}

\providecommand{\lMax}{\ell_\mathrm{max}}
\providecommand{\omMax}{\omega_\mathrm{max}}
\providecommand{\g}{{\mathfrak{g}}}
\providecommand{\h}{{\mathfrak{h}}}

\DeclareMathOperator{\DST}{DST}

\providecommand{\df}{{\rm d}}

\begin{document}

\title{\boldmath A new basis for Hamiltonian \texorpdfstring{\sutwo}{SU(2)} simulations}

\author{Christian W. Bauer}
\email{cwbauer@lbl.gov}
\affiliation{Physics Division, Lawrence Berkeley National Laboratory, Berkeley, CA 94720, USA}
\affiliation{Department of Physics, University of California, Berkeley, Berkeley, CA 94720}

\author{Irian D'Andrea}
\email{irian@berkeley.edu}
\affiliation{Physics Division, Lawrence Berkeley National Laboratory, Berkeley, CA 94720, USA}
\affiliation{Department of Physics, University of California, Berkeley, Berkeley, CA 94720}

\author{Marat Freytsis}
\email{mf823@cornell.edu}
\affiliation{LEPP, Department of Physics, Cornell University, Ithaca, NY 14853, USA}

\author{Dorota M. Grabowska}
\email{grabow@uw.edu}
\affiliation{InQubator for Quantum Simulation (IQuS), Department of Physics, University of Washington, Seattle, WA 98195}

\preprint{IQuS@UW-21-058}
\date{\today}

\begin{abstract}
Due to rapidly improving quantum computing hardware, Hamiltonian simulations of relativistic lattice field theories have seen a resurgence of attention. 
This computational tool requires turning the formally infinite-dimensional Hilbert space of the full theory into a finite-dimensional one. 
For gauge theories, a widely-used basis for the Hilbert space relies on the representations induced by the underlying gauge group, with a truncation that keeps only a set of the lowest dimensional representations. 
This works well at large bare gauge coupling, but becomes less efficient at small coupling, which is required for the continuum limit of the lattice theory.  
In this work, we develop a new basis suitable for the simulation of an \sutwo lattice gauge theory in the maximal tree gauge. 
In particular, we show how to perform a Hamiltonian truncation so that the eigenvalues of both the magnetic and electric gauge-fixed Hamiltonian are mostly preserved, which allows for this basis to be used at all values of the coupling. 
Little prior knowledge is assumed, so this may also be used as an introduction to the subject of Hamiltonian formulations of lattice gauge theories.

\end{abstract}

\maketitle
\section{Introduction}

Hamiltonian simulations of relativistic lattice field theories have recently received a lot of attention in the nuclear and high energy physics community~\cite{Mil:2019pbt,Yang:2020yer,Zhou:2021kdl,Gorg:2018xyc,Stannigel:2013zka,Zohar:2012xf,Mezzacapo:2015bra,Bazavov:2015kka,Yang:2016hjn,Gonzalez-Cuadra:2017lvz,Davoudi:2019bhy,Surace:2019dtp,Luo:2019vmi,Banerjee:2012pg,Ott:2020ycj,Dasgupta:2020itb,Osborne:2022jxq,Tagliacozzo:2012df,Zohar:2013zla,Hauke:2013jga,Kuhn:2014rha,Kasper:2015cca,Zohar:2015hwa,Alexandru:2019nsa,Klco:2019evd,Davoudi:2020yln,Haase:2020kaj,Kreshchuk:2020dla,Ciavarella:2021nmj,Buser:2020cvn,Ciavarella:2021nmj,Bauer:2021gek,Byrnes:2005qx,Grabowska:2022uos,Kane:2022ejm,Anishetty:2009nh,Banuls:2017ena,Kaplan:2018vnj,Zohar:2019ygc,Raychowdhury:2018osk,Raychowdhury:2019iki,Stryker:2021asy,Davoudi:2022xmb,Barata:2020jtq,Shaw:2020udc,Kan:2021xfc,Lamm:2019bik,ARahman:2022tkr,ARahman:2021ktn,Martinez:2016yna,Nguyen:2021hyk,Klco:2018kyo,Kokail:2018eiw,Nagano:2023uaq}. The basic tool is Hamiltonian lattice field theory which was pioneered in~\cite{Kogut:1974ag,Banks:1975gq}.
For recent reviews on this topic, please see~\cite{Banuls:2019bmf,Kasper:2020akk,Aidelsburger:2021mia,Klco:2021lap,Bauer:2022hpo}.
A major reason for the resurgence in interest is in large part due to recent advances in quantum computation that should make such calculations feasible in the future.
In particular, it has been shown rigorously that a quantum computer allows all aspects of a scalar field theory to be simulated with resources that scale logarithmically with the dimension of the truncated Hilbert space~\cite{Jordan:2011ci,Jordan:2012xnu}, providing exponential improvement over the polynomial resource scaling of the best known algorithms on classical computers.

In general, the Hamiltonian of a pure gauge theory can be written in terms of two components, describing the contributions to the energy from the electric and magnetic fields
\begin{equation}
    H \sim H_E + H_B \,.
\end{equation}
After appropriate field redefinitions, the electric Hamiltonian is conventionally proportional to $g^2$, where $g$ denotes the (bare) spatial coupling constant, while $H_B$ is proportional to $g^{-2}$. 
Therefore the electric Hamiltonian dominates at large coupling, while the magnetic Hamiltonian dominates at small coupling. 
The electric and magnetic Hamiltonians do not commute with one another, which implies that one cannot find a basis for the Hilbert space in which both the electric and magnetic Hamiltonians are simultaneously diagonal.

As will be discussed in much more detail, there are two main classes of bases that can be chosen for the Hilbert space of a lattice gauge theory. 
The first class of bases is called the irrep basis,  where each state is labelled by an irreducible representation (irrep) of the gauge group. 
In this basis the electric Hamiltonian is diagonal. 
One major advantage of the irrep basis is that the constraint from Gauss's law is expressed simply, since it is directly related to the combination of representations of links emanating from a given vertex.
At large coupling, where the electric Hamiltonian dominates, the lowest lying states are dominated by the lowest lying representations of the gauge group. For an \sutwo gauge theory this implies that the irrep basis can easily be truncated by removing representations with $j > j_\mathrm{max}$ for some appropriately chosen cutoff $j_\mathrm{max}$.
The irrep basis underlies most recent formulations of \sutwo and other non-Abelian gauge theories to date~\cite{Raychowdhury2023,Mathur:2004kr,mathur2006loop,mathur2007loop,Klco:2019evd,Davoudi:2020yln,Ciavarella:2021nmj,Byrnes:2005qx,Anishetty:2009nh,Banuls:2017ena,Raychowdhury:2018osk,Raychowdhury:2019iki}.

At small coupling, however, the magnetic Hamiltonian dominates, and the lowest lying eigenstates are superpositions of a large number of irreps. 
This makes the irrep basis and its variants inefficient at small couplings, since a large number of states have to be kept in the truncation.
In this limit, it is more convenient to work in the group element basis, where each basis state at a given link corresponds to a particular group element of \sutwo at that link. 
In this basis the magnetic Hamiltonian will be diagonal. 
It turns out that is quite challenging to formulate continuous gauge theories in the group element basis. For a \uone gauge theory a formulation at weak coupling was proposed in~\cite{Haase:2020kaj}, and a formulation that works at all values of the coupling was recently developed~\cite{Bauer:2021gek}.
For non-Abelian theories, most formulations of the group element basis have focused on working with discrete subgroups~\cite{Zohar:2014qma, Alexandru:2019nsa,Lamm:2019bik, Lamm2022,Lamm2020}.

Developing formulations of the group element basis is of key importance for the following reason.
The coupling constant $g$ appearing in lattice formulations is a bare coupling constant, meaning it is not the one that appears in continuum formulations at low energies. 
Rather it must be tuned as a function of the lattice spacing so as to reproduce some long distance physical measurement (often the mass of some particle or bound state). 
The non-Abelian theories of interest (especially quantum chromodynamics) are asymptotically free, meaning the bare coupling constant becomes small at small distances. 
For this reason it is important to be able to efficiently simulate such theories at weak bare coupling, even if the theory we ultimately seek is strongly coupled in the infrared.

This work develops a formulation of \sutwo lattice gauge theory with resource requirements that do not have a strong dependence on the value of the coupling. 
It will be shown how one can choose a particular basis (which we call the mixed basis) for the Hilbert space of an \sutwo lattice gauge theory which allows for a truncation that is independent of the relative size of the electric and magnetic Hamiltonian.
This work is organized as follows: In \cref{sec:basics} we state some basic properties of \sutwo gauge theories and establish the notation used. 
In \cref{sec:bases_and_trucation} the electric and magnetic bases will be described in more detail. 
We will also introduce a coordinate system for the magnetic basis which will be useful for developing the mixed basis later.
In \cref{sec:singlePlaquette} we will study the simplest system consisting of a $2\times 2$ lattice with open boundary conditions, which contains only a single plaquette. We will use this to introduce the mixed basis, as well as an associated truncation scheme. Some numerical results are presented.
In \cref{sec:multiPlaquette} these ideas will be extended to a general lattice, before we conclude in \cref{sec:conclusions}. 

\section{\boldmath Basics of \texorpdfstring{\sutwo}{SU(2)} Lattice Gauge Theory} 
\label{sec:basics}
In this section we will introduce the basics of pure \sutwo lattice gauge theory (LGT), and introduce the notation and conventions used in this work.
Little prior knowledge is assumed, and some readers may wish to skip or skim certain sections in order to get to the main results of this paper.

For a particle physicist interested in continuum physics, LGTs are regularized field theories with an ultraviolet cutoff $\Lambda \sim 1/a$ set by the lattice spacing and an infrared cutoff $\sim 1/L$ set by the size of the lattice. 
One always has in mind the double limit $a \to 0$, $L \to \infty$, while simultaneously tuning the parameters of the theory in such a way as to reproduce some physical observables (such as the mass of some bound states), thereby hopefully reproducing the continuum theory. 
As the lattice spacing and lattice volume simultaneously regulate both the action and the measure of the path integral, the theory is free of divergences and therefore LGTs are well-defined outside of perturbation theory even at strong coupling. 
This makes them of particular interest for studying the confinement of quarks and other strongly coupled systems.

Most frequently this is done in the Euclidean path integral approach via Markov-chain Monte Carlo (MCMC) methods, see~\cite{FlavourLatticeAveragingGroupFLAG:2021npn,Davoudi:2022bnl,USQCD:2022mmc,Davoudi:2020ngi} for recent reviews. 
This paper discusses the Hamiltonian approach to the same theory.

\subsection{Classical Yang--Mills Theory}

The continuum field theory of interest is pure Yang--Mills, containing only self-interacting gauge bosons. 
When it does not complicate the presentation, statements will be formulated to also be applicable to gauge groups other than \sutwo.

The theory rests on the structure of a compact Lie group $G$, whose underlying Lie algebra has Hermitian generators $T^a$ satisfying
\begin{equation}
    [T^a, T^b] = if^{abc} T^c
\end{equation}
for totally antisymmetric structure constants $f^{abc}$. 
For \sutwo, the structure constants are $f^{abc} = \epsilon^{abc}$. 
We normalize the generators of the defining representation so that $\Tr(T^a T^b) = \frac{1}{2} \delta_{ab}$. 
For \sutwo, they can be chosen to be half the Pauli matrices $T^a = \frac{!}{2}\sigma^a$, while for \suthree they would be the Gell-Mann matrices.

The gauge bosons are represented by a collection of spin-1 fields $A_\mu^a$ labelled by index $a = 1,\dotsc, \dim(G)$. 
The fields are often packaged into a Lie algebra-valued field
\begin{equation}
    A_\mu = A_\mu^a T^a \,.
\end{equation}
The mathematical function of the gauge field is one of a connection, and it is properly described using the language of fiber bundles. 
For our purposes the important point is that the theory contains a local gauge redundancy with parallel transport provided by the field $A_\mu$. 
Fields participating in the gauge symmetry enter into the action in such a way as to be invariant under independent color rotations at every point in spacetime.

Let $\psi(x)$ be a field at position $x$ in representation $r$ of the gauge group. 
It can be thought of as a vector of dimension $\dim(r)$, with the two cases of most interest being the defining representation (the spin-1/2 representation for \sutwo) and the adjoint representation (having the same dimension as the Lie group itself). 
We demand that $\psi(x)$ enters into the action in such a way as to be invariant under the transformation
\begin{equation}
    \psi(x)^j \longmapsto D^r(\Omega(x))^i_j \psi^j(x) 
\end{equation}
for all group-valued functions $\Omega(x) \in G$. 
The set of all such functions $\Omega$ is the \emph{gauge group} of the theory, though we may sometimes also refer to the group $G$ as the gauge group\footnote{Note that $G$ is much smaller than $\Omega$ as it is only a single copy, and is also often referred to as the structure group}. 
The object $D^r(\Omega)^i_j$ is the group element $\Omega$ in representation $r$, \ie, a $\dim(r) \times \dim(r)$ matrix.

In taking the derivative of the field $\psi$, one must compare its values at two nearby points 
\begin{equation}
  \lim_{a \to 0} \frac{\psi(x+a) - \psi(x)}{a} \,,
\end{equation}
and that this object is not gauge covariant, as one may independently rotate the two values in color space.
In order to construct a kinetic term, one needs a way of comparing the values of color-charged objects at different points. 
This is provided by the connection, and the object
\begin{align}
    \mathcal{D}_\mu \psi(x) = \partial_\mu \psi(x) - i A_\mu(x) \psi(x)
\end{align}
is gauge covariant (transforming in the same way as $\psi$ itself) provided the connection $A_\mu(x)$ also undergoes gauge transformation 
\begin{equation}\label{eq:connection_gauge_transform}
    A_\mu(x) \longmapsto \Omega(x) A_\mu \Omega^{-1}(x) + i \Omega(x)\, \partial_\mu \Omega^{-1}(x) \,.
\end{equation}
The transformation properties of $A_\mu$ are natural in that they lead to parallel transport, in the following sense.

\begin{figure}[h]
\centering
\includegraphics[width=0.24\textwidth]{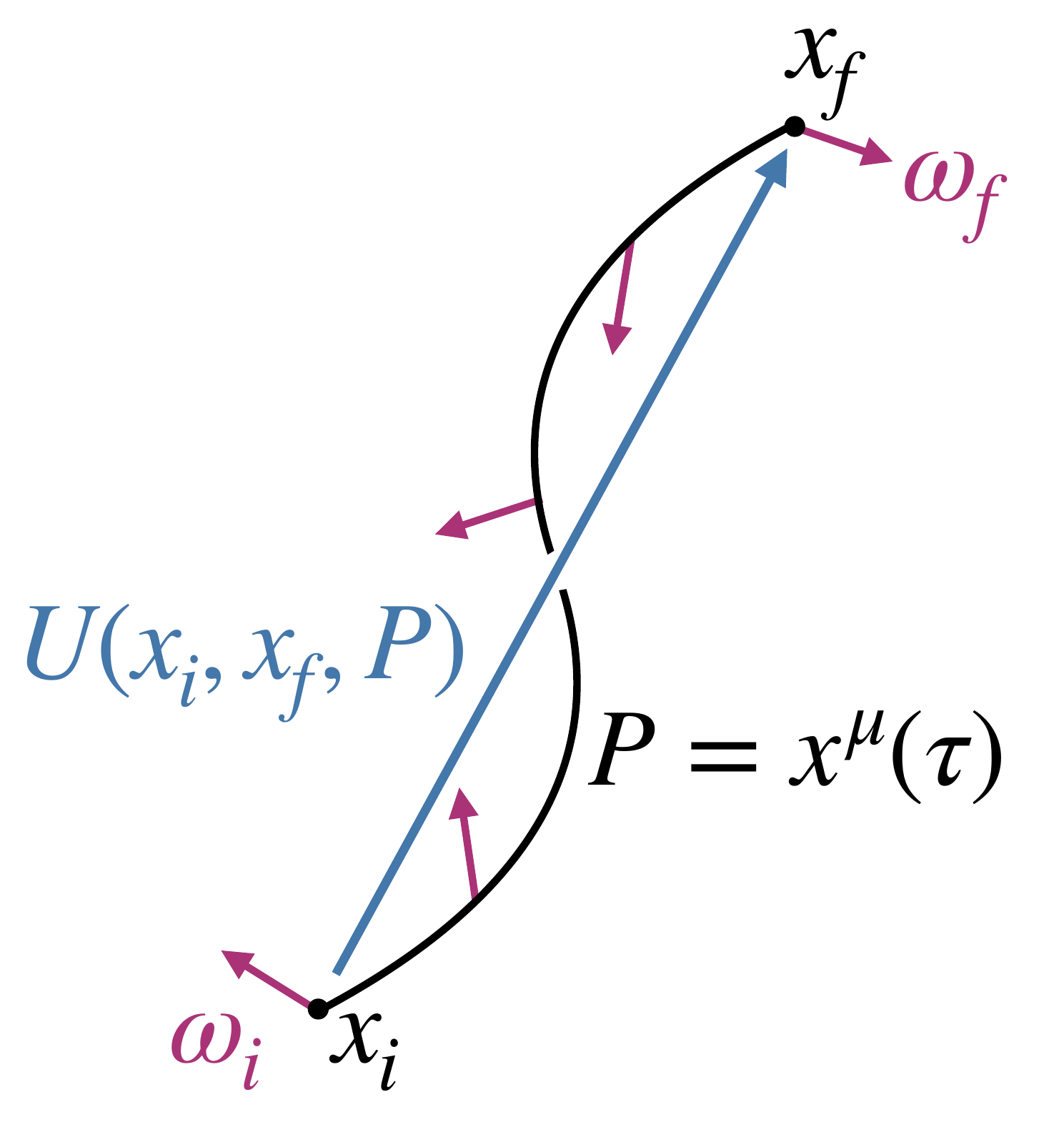}
\caption{A color charge $w$ (indicated by the purple arrows) being transported along world-line $P = x^\mu(\tau)$. The Wilson line $U(x_i, x_f, P)$ transports the color charge as indicated in \cref{eq:w_transport}.}
\label{fig:w_transport}
\end{figure}

Consider a color-charged point particle in the representation $r$, which can also be thought of as having a color vector $w$ of dimension $\dim(r)$ attached to it, and is in fact the semi-classical limit of a single excitation of the field $psi(x)$ above.
We imagine that the particle follows some prescribed path along worldline $x^\mu(\tau)$ in the presence of a background connection $A_\mu(x)$. 
Free of external forces, the particle's color charge $w(x)$ satisfies the parallel transport equation
\begin{equation}\label{eq:parallel_transport}
    \dv{w(x(\tau))}{\tau} = i \dv{x^\mu(\tau)}{\tau} A_{\mu}(x(\tau)) \,  w(x(\tau)) \,.
\end{equation}
Note that both $w(x)$ and $A_\mu(x)$ depend on the representation $r$.
This states that the covariant derivative of $w(x(\tau))$ along the direction of its own velocity vector $\dot{x}^\mu(\tau)$ is zero. 
Since $A_\mu$ is Hermitian, the form of the right-hand side of the equation guarantees that the magnitude of $w$ will be constant. 
Consider the path $P$ beginning at $x_i \equiv x(\tau_i)$ and ending at $x_f \equiv x(\tau_f)$. 
The solution to the differential equation then takes the form
\begin{equation}\label{eq:w_transport}
    w(x_f) = U(x_f,x_i;P) w(x_i)
\end{equation}
for some unitary matrix $U$ depending on the starting pointing, the ending point, and the path taken. 
For an illustration of these concepts, see \cref{fig:w_transport}.

\Cref{eq:parallel_transport} can be straightforwardly integrated to arrive at
\begin{equation}
\begin{split}
U(x_f,x_i;P) &= \mathcal{P}\exp( i\int_{\tau_i}^{\tau_f} \dd{\tau}\dv{x^\mu}{\tau} A_{\mu}(x(\tau)) ) \\
&= \mathcal{P}\exp( i\int_{x_i}^{x_f} \dd{x^\mu} A_{\mu}(x) )
\end{split}
\end{equation}
where $\mathcal{P}\exp$ indicates a path-ordered exponential, such that in the Taylor expansion of the exponential, any $A_\mu(\tau)$ at earlier times on the path are placed to the right of those at later times. 
The connection $A_\mu$ appearing in this equation may be put into any representation $A_\mu = A_\mu^a T^{ra}$, in which case the resulting object transports charges in representation $r$. 
The object $U(x_i,x_f;P)$ is called a \emph{Wilson line}, and on the lattice this becomes the basic object of the theory more-so than the connection itself.
Using \cref{eq:connection_gauge_transform} and the properties of path-ordering, one finds the gauge transformation property
\begin{equation}\label{eq:wilson_line_gauge_trans}
    U(x_f, x_i; P) \longmapsto \Omega(x_f) U(x_f, x_i; P) \Omega(x_i)^\dagger \,.
\end{equation}
If the Wilson line is in a representation other than the defining representation, $\Omega$ is replaced with the appropriate $D(\Omega)$ matrix. 
Note that $D^{\mathrm{defining}}(\Omega)=\Omega $.

As \cref{eq:wilson_line_gauge_trans} shows, the Wilson line only takes on a gauge-independent value, and is physically meaningful, when taken around a closed loop $C$.
For instance, in the Abelian case, the phase $U(C) = \exp(i\oint A)$ captures the interference in the Aharonov--Bohm effect. 
If we wish to construct local gauge invariant objects, we are then led to consider parallel transport around an infinitesimal loop.

Consider a very small parallelogram with sides $\xi^\mu$ and $\zeta^\nu$. 
We wish to determine the effect of parallel transporting charge $w$ around the closed loop $C$ in the limit of the side lengths $\xi, \zeta \to 0$. 
In this limit we can write the answer as $[U(C)w-w] = -i F_{\mu\nu} \xi^\mu \zeta^\nu w$. 
By Taylor expanding the path-ordered exponential to second order, one finds that $F_{\mu\nu}$ is the standard field strength tensor
\begin{equation}\label{eq:F_def}
    F_{\mu\nu} = \partial_\mu A_\nu - \partial_\nu A_\mu - i \comm{A_\mu}{A_\nu} \,.
\end{equation}
Like the connection, this object is Lie-algebra-valued and may be expanded as $F_{\mu\nu} = F_{\mu\nu}^a T^a$. 
Unlike the connection, this object is gauge-covariant, transforming like
\begin{equation}
F_{\mu\nu}(x) \longmapsto \Omega(x) F_{\mu\nu}(x) \Omega(x)^\dagger \,.
\end{equation}
Its trace is therefore a local gauge invariant object.

Given these results, we are now in a position to write down the standard Yang--Mills action (without a theta term, as we are not concerned with topological effects here). 
The simplest local gauge invariant action formed out of the gauge boson field is
\begin{equation}\label{eq:YM_action}
    S_\mathrm{YM} = -\frac{1}{2g^2} \int \dd[4]{x} \Tr F_{\mu\nu} F^{\mu\nu} \,,
\end{equation}
which is the standard Yang--Mills action. 
The placement of the coupling constant $g$ and the factor of 1/2 are conventional and may be absorbed into $F_{\mu\nu}$ through field redefinition, though we prefer this normalization since it reproduces the standard normalization for the lattice case.

Since we are ultimately interested in the canonical quantization of this theory, we turn to the Hamiltonian formulation. 
The canonical momenta are
\begin{align}\label{eq:can_momenta}
\pi^0 &= \pdv{\mathcal{L}}{\dot{A_0}} = 0 \\
\pi^i &= \pdv{\mathcal{L}}{\dot{A_i}} = -\frac{1}{g^2} F^{0i} \equiv \frac{1}{g^2}E^i \,,
\end{align}
where we have defined the chromo-electric field $\bm{E} = \bm{E}^a T^a$. 
We see that the component $A_0$ is non-dynamical, as in the Abelian case. 
Similarly defining the chromo-magnetic field to be $\epsilon_{ijk}B^a_k = F^a_{ij}$, the canonical Hamiltonian is
\begin{align} \label{eq:continuum_ham}
\mathcal{H} &= \pi^{ia} \dot{A}^a_i - \mathcal{L} \nonumber\\ 
&= \frac{1}{2g^2} \pqty{ E^{ia}E^{ia} + B^{ia}B^{ia}- 2 A_0^a (\partial_i E^{ia} + f^{abc}A_{i}^b E^{ic}) }  \nonumber\\
&= g^2 \Tr \bm{\pi}^2 + \frac{1}{g^2}\Tr \bm{B}^2 - \frac{1}{g^2} A_0^a (\mathcal{D}_i E^i)^a \,,
\end{align}
where the covariant divergence of the electric field $(\mathcal{D}_i E^{i})$ is given by 
\begin{equation}
  (\mathcal{D}_i E^{i}) \equiv \partial_i E^{i} - i \comm{A_i}{E^i} \,,
\end{equation}
and in the last step we expressed the Hamiltonian in terms of the canonical momenta $\pi^i$ to illustrate how the coupling constant will appear in the lattice theory.

When working with the Hamiltonian it is convenient to work in the Weyl gauge $A_0=0$, sometimes also referred to as the Hamiltonian gauge or temporal gauge. 
A result of this choice of gauge is that the covariant time derivative is an ordinary time derivative, such that both sides of the Schrodinger equation $i \partial_t \ket{\psi} = H \ket{\psi}$ are gauge invariant (and in particular the Hamiltonian is gauge invariant).
This is an incomplete gauge-fixing condition, as arbitrary residual gauge transformations $\Omega(\bm{x})$, which have only spatial dependence, are still allowed.

Another consequence of this choice of gauge is that Gauss's law becomes a supplementary constraint, rather than a Hamiltonian equation of motion. 
This is apparent if one examines the Hamiltonian of \cref{eq:continuum_ham}, in which $A_0$ appears as a Lagrange multiplier enforcing Gauss's law $\mathcal{D}_i E^i=0$ in the absence of charged matter. 
Then, when $A_0=0$ is imposed at the outset, in the quantized theory Gauss's law must be imposed as a condition on the states of the Hilbert space. 
This will be discussed in more detail later for the lattice theory.

\subsection{The Kogut--Susskind Hamiltonian}
\label{sec:basics_KG}

We now review the Hamiltonian formulation of lattice gauge theory developed by Kogut and Susskind \cite{Kogut:1974ag,Banks:1975gq}. 
Again, readers familiar with this subject may wish to skip this section, although the literature on this subject is filled with differing conventions so it may be useful in any case. 
We follow here mostly the discussion in~\cite{Zohar:2014qma} and \cite{Zohar:2021nyc}.

In the Hamiltonian formulation of a lattice gauge theory, time remains a continuous variable while the previously continuous spatial degrees of freedom are replaced by a discrete set associated with a lattice with spacing $a$. 
We will take the spatial lattice to be hypercubic, with sites $n$ and links $\ell = \ell(n,e_i)$ joining adjacent sites $n$ and $n+e_i$. 
Here $e_i$ is a unit vector along the $i$th positive direction for $i = 1,\dotsc,d$.

Recall that the basic function of the gauge field is to transport color charge. 
Since color charges reside only on sites $n$, the only necessary quantities are the Wilson lines $U(n_f,n_i;P)$ joining adjacent sites, and products thereof. 
It will prove inconvenient to attempt to work with the potential $A_\mu$, and we instead work directly with the Wilson lines $U(n, e_i)\equiv U(n,n+e_i;P_\ell)$ regarded as a single degree of freedom through coarse-graining. Here $P_\ell$ is the straight line along the link connecting the two adjacent points.

We will now simultaneously develop the latticization and quantization of this theory, starting with the Hilbert space at a single link of the lattice. 
As just mentioned, the single degree of freedom on link $\ell$ is the Wilson line $U_\ell \in G$, hereafter also referred to simply as the link variable, or link operator after quantization. 
The Hilbert space of a single link is $L^2(G,\dd{\g})$, the space of square-integrable wave functions over the Lie group $G$ with respect to the Haar measure $\dd{\g}$.
One way of describing this space is through the so-called group element basis, which are states of definite value on \sutwo and can be thought of as position eigenstates. 
A general state may be expanded in the group element basis as
\begin{equation}
\ket{\Psi} = \int\dd{\g} \Psi(\g) \ket{\g} \,.
\end{equation}
Group element states $\ket{\g}$ are orthogonal,
\begin{equation}
\braket{\g}{\h} = \delta_{\g\h} \,,
\end{equation}
where the delta function is a distribution satisfying
\begin{equation}
\int\dd{\h} \ket{\h} \delta_{\g\h} = \ket{\g} \,.
\end{equation}
The link $\ell(n,e_i)$ has an orientation inherited from the lattice positive direction $+e_i$. 
We refer to the oriented start of the link also as the \emph{left} side of the link, and the oriented end of the link as the \emph{right} side of the link. 

From \cref{eq:wilson_line_gauge_trans} gauge transformations enacted at sites $n$ and $n+e_i$ will affect the state of the link via left and right group composition, respectively. 
We then define left and right translation operators, parameterized by a group element $\g$, which act on a basis state $\ket{\h}$ at each link $\ell$ via left and right group composition as
\begin{equation}
\label{eq:link_trafo}
\hat{\Theta}_{L\g} \ket{\h} = \ket{\g^{-1}\h} \qc\hat{\Theta}_{R\g} \ket{\h} = \ket{\h\g^{-1}} \,.
\end{equation}
For a compact Lie group $G$, these translation operators can always be expressed as
\begin{equation}\label{eq:E_def}
\hat{\Theta}_{L\g} = e^{i \phi^a(\g) \hat E_L^a} \qc\hat{\Theta}_{R\g} = e^{i \phi^a(\g) \hat E_R^a} \,,
\end{equation}
where $\phi^a(\g)$ is a set of $\dim(G)$ parameters, called normal coordinates, or exponential coordinates, specifying the group element $\g$, and the Hermitian operators $\hat E_L$ and $\hat E_R$ furnish two independent Lie algebras of the group
\begin{equation}
\label{eq:EEcommutators}
\begin{split}
\comm{\hat{E}_L^a}{\hat{E}_L^b} &= -i f^{abc} \hat{E}_L^c \\
\comm{\hat{E}_R^a}{\hat{E}_R^b} &= i f^{abc} \hat{E}_R^c \\
\comm{\hat{E}_L^a}{\hat{E}_R^b} &= 0 \,.
\end{split}
\end{equation}
These commutators follow from \cref{eq:link_trafo,eq:E_def}. 
Note the minus sign in the commutator of the left electric field, often referred to as an \emph{anomalous} commutator. 
The minus sign can be removed by redefining the electric operators, \eg through $\hat{E}_L^a \to -\hat{E}_L^a$, as is done by some authors. 
We prefer the convention above, as it causes the left and right electric operators to be related by parallel transport. 
As we will see, these operators have the physical meaning of measuring the chromo-electric field at the left and right ends of the link, respectively. 
That they are not equal to each other reflects the fact that gauge fields carry color charge and cause the electric field to rotate upon traversing the link.

Note that the words ``left'' and ``right'' have been overloaded to have two meanings: one refers to \emph{group composition} which may proceed either by left-multiplication or right-multiplication, and the other refers to the two sides of a link. 
Our conventions have been chosen in such a way as to refer to the same thing; the left electric field is associated with the left side of the link, and it generates left translation.

In the representation theory of compact Lie groups, the irreducible representations (irreps) may always be taken to be unitary matrices $D^r_{m m'}$ with dimension $\dim(r) \times \dim(r)$.
For the case of \sutwo, these are the spin-$j$ representations with $j=0,\frac{1}{2},1, \dotsc$ and dimension $2j+1$, which we will denote $D^j_{m m'}(\g)$. 
We define the (spin-$j$) unitary link operators by
\begin{equation}
\label{eq:def_link_op}
\hat{U}^j_{mm'}(\ell) \equiv \int\dd{\g_\ell} D^j_{{mm'}}(\g_\ell) \op{\g_\ell} \,,
\end{equation}
where $\ell$ is used to denote the specific link with which this operator is associated.
These operators are diagonal in the group element basis and can be thought of as position operators. 
Since they are all simultaneously diagonal, all link operators commute with one another. 
Note that while these operators commute,
\begin{equation}
  \comm{\hat{U}^j_{mn}(\ell)}{\hat{U}^k_{rs}(\ell')} = 0 \,,
\end{equation}
care must be taken about their ordering when summing over the $mn$ indices, as 
\begin{align}
\hat U^j_{mn}(\ell)\hat U^j_{nk}(\ell') \neq \hat U^j_{mn}(\ell')\hat U^j_{nk}(\ell)
\,,
\end{align}
except for $m = k$. 
This is of particular importance when constructing plaquette operators.

Using \cref{eq:link_trafo,eq:def_link_op} one can show that these operators transform under left and right translation via
\begin{equation}\label{eq:U_translation}
  \begin{split}
    \hat{\Theta}_{L\g} \hat{U}^j_{mn} \hat{\Theta}^\dagger_{L\g}
      &= D^j_{{m}{m'}}(\g) \hat{U}^j_{m'n} \,, \\
    \hat{\Theta}_{R\g} \hat{U}^j_{mn} \hat{\Theta}^\dagger_{R\g}
      &= \hat{U}^j_{mn'} D^j_{n'n}(\g) \,.
  \end{split}
\end{equation}
This implies the commutation relations between the link operators and electric operators
\begin{equation}\label{eq:EUcommutators}
  \begin{split}
    \comm{\hat{E}_L^a}{\hat{U}^j_{mn}} &= T^{ja}_{mm'}\hat U^j_{m'n} \, \\
    \comm{\hat{E}_R^a}{\hat{U}^j_{mn}} &= \hat{U}^j_{mn'}T^{ja}_{n'n} \,,
  \end{split}
\end{equation}
where the $T^{ja}$ denote the generators of the Lie group in the representation $j$, and $j$ is not summed over. 
When no representation label is written, the defining one is assumed.

The left and right electric operators are related to each other by parallel transport
\begin{equation}
    \hat{E}_L^a = \hat{U}^{1}_{ab} \hat{E}_R^b =  \hat{E}_R^b \hat{U}^{1}_{ab} \,,
\end{equation}
where the $j=1$ representation is the adjoint one so the indices $mn$ are replaced with group indices $ab$. 
Note that $\hat{E}^a_R$ and $\hat{U}^1_{bc}$ don't commute in general, but a contraction in the group indices will cause the commutator to vanish, \ie,
\begin{equation}
  \comm{\hat E^a_R}{\hat U^1_{ab}} = 0 \,.
\end{equation}

The link operator in the adjoint representation can be expressed as
\begin{equation}\label{eq:adjoing_def}
    \hat{U}^1_{ab} \equiv 2\Tr( \hat{U}^\dagger T^a \hat{U} T^b ) \,,
\end{equation}
where $\hat{U}$ is the link operator in the defining representation. 
From this, and the unitarity of the link operator matrix $(\hat{U}^\dagger)_{nm'} \hat{U}_{m'm} = \delta_{nm} \mathds{1}$, it is easy to show that the adjoint link operator is an orthogonal matrix of operators
\begin{equation}
    \hat{U}^{1}_{ac} \hat{U}^1_{bc} = \delta_{ab}\mathds{1} \,.
\end{equation}
Then because the left and right electric operators are related by parallel transport, their quadratic Casimirs are equal to each other as operators
\begin{equation}\label{eq:Casimir}
    \hat{E}^a_L \hat{E}^a_L = \hat{E}^a_R \hat{E}^a_R \equiv \hat{E}^2 \,.
\end{equation}

We are now in a position to describe the Hilbert space in terms of a complete set of commuting electric operators, which we will take to be $\hat{E}^z_L$, $\hat{E}^z_R$, and $\hat{E}^2$. 
The Lie algebra commutation relations among the electric operators \cref{eq:EEcommutators} imply that the simultaneous eigenstates of $\hat{E}^2$ and $\hat{E}_L^z$ form an angular momentum multiplet structure $\ket{jm}$, as do the simultaneous eigenstates of $\hat{E}^2$ and $\hat{E}_R^z$, $\ket{j'm'}$. 
However from \cref{eq:Casimir} we know that they share a quadratic Casimir $j=j'$ and the states can therefore be written as
\begin{equation}
    \ket{j{m_L}{m_R}}\, 
\end{equation}
with $j, m_L, m_R$ being the relevant quantum numbers. 
The situation is mathematically equivalent to the quantum mechanics of a rigid body with three non-vanishing moments of inertia, in which $j$ is the total angular momentum, $m_L$ is the $z$-component of the angular momentum with respect to space-fixed axes, $m_R$ is the $z$-component of the angular momentum with respect to body-fixed axes, and $\hat{U}^1_{ab}$ is the rotation matrix mapping space axes to body axes.
The quantum number $j=0,\frac{1}{2},1,\frac{3}{2},\dotsc$ takes on half-integer values and $m_L,m_R$ each independently range from $-j$ to $+j$ in integer steps. For a fixed half-integer $j$, there are then $(2j+1)^2$ states that have total ``angular momentum'' $j$.

The overlap of these states with the group element basis states is given again by the representation matrix
\begin{equation}\label{eq:FT_SU2}
    \braket{\g}{j m_L m_R} = \sqrt{\frac{\dim(j)}{|G|}} D^j_{{m_L}{m_R}}(\g) \,,
\end{equation}
which are the matrix elements of the non-Abelian generalization of the Fourier transform, moving between group elements (position) and irreducible representations (momenta). 
Here $\dim(j) = 2j+1$ is the dimension of the spin $j$ representation and $|G| = \int\dd{\g}$ is the volume of the group when integrated over the Haar measure, which is given by to $|G| = 16\pi^2$ for \sutwo.
As already mentioned, the irrep basis states are eigenstates of a maximal commuting set of the electric operators
\begin{equation}\label{eq:Eop_electric_basis}
  \begin{split}
    \hat{E}^2 \ket{j m_L m_R} &= j(j+1) \ket{j m_L m_R} \\
    \hat{E}^z_L \ket{j m_L m_R} &= m_L \ket{j m_L m_R} \\
    \hat{E}^z_R \ket{j m_L m_R} &= m_R \ket{j m_L m_R} \,.
  \end{split}
\end{equation}

We now return to the construction of the lattice Hamiltonian. 
On each link of the lattice one places a copy of the one-link Hilbert space just described, with operators labelled by the link $\ell(n,e_i)$ to which they belong, \eg, $\hat{U}_{\ell} = \hat{U}(n,e_i)$. 
Operators on different links commute with each other.

\begin{figure}[b]
\centering
\includegraphics[width=0.45\textwidth]{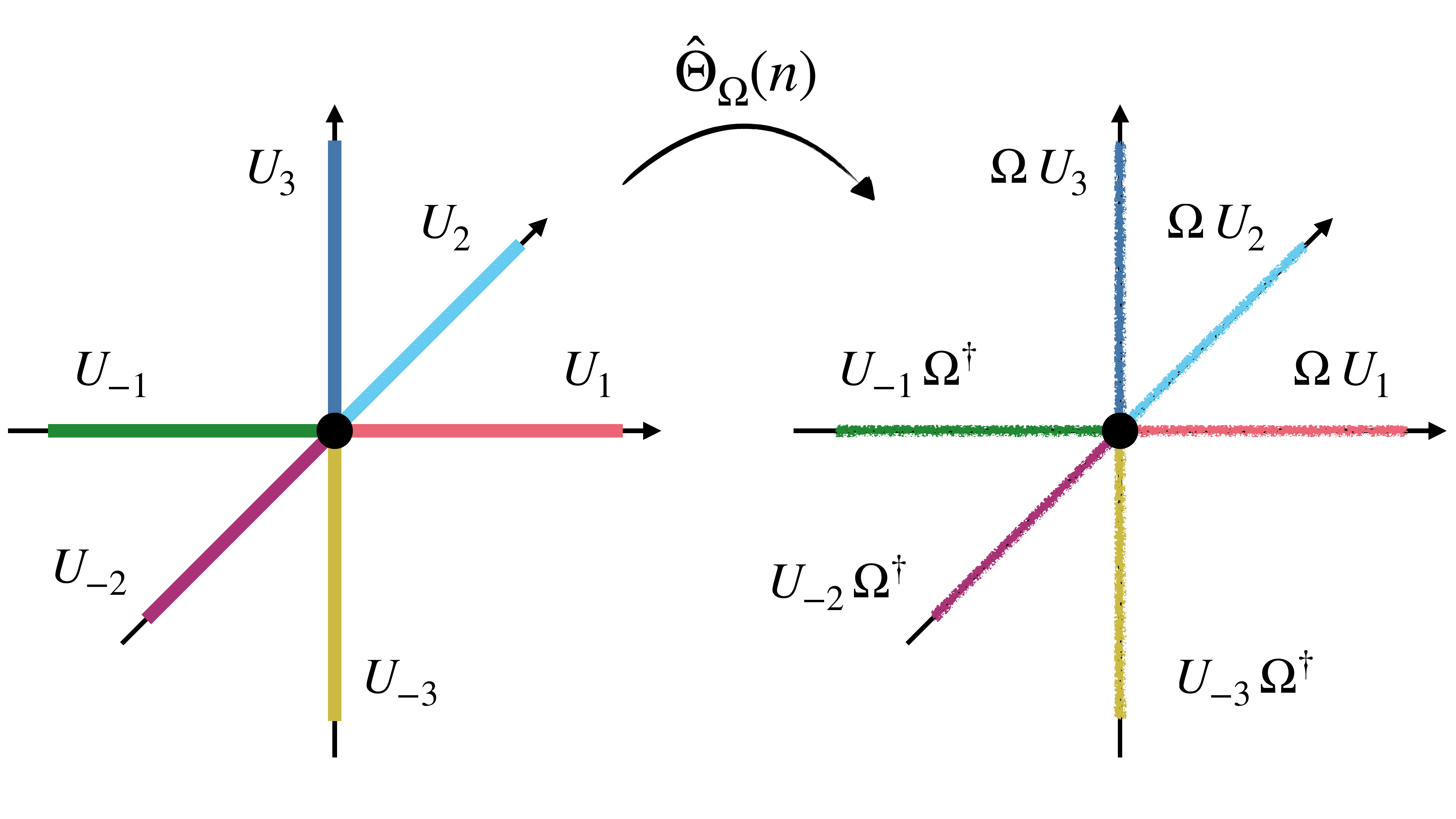}
\caption{The action of the gauge transformation $\hat{\Theta}_\Omega(n)$ at site $n$, viewed passively.}
\label{fig:gauge_trans}
\end{figure}

Consider a gauge transformation $\Omega(n)$, taken now as an independent color rotation at each lattice site rather than at every location in spacetime. 
Note that as discussed previously, we are already working in Weyl gauge $A_0=0$, so the gauge transformations only have spatial dependence. 
Consider for a moment a gauge transformation affecting only a single lattice site $n$, keeping in mind that a general gauge transformation can be made by combining such one-site transformations. 
Let $\Omega=\Omega(n)$ be the transformation at site $n$.
The effect of the gauge transformation on the local link operators is shown in \cref{fig:gauge_trans}.
In words, all link operators extending from site $n$ in a positive direction $\hat{U}(n,e_i)$ are affected at their left endpoint by a gauge transformation as in \cref{eq:wilson_line_gauge_trans}, which with the help of \cref{eq:U_translation} can be expressed as
\begin{equation}\label{eq:U_left_trans}
    \hat{U}^j_{mn}(n,e_i) \longmapsto D^j_{{m}{m'}}(\Omega) \hat{U}^j_{m'n}
      = \hat{\Theta}_{L\Omega} \hat{U}^j_{mn} \hat{\Theta}_{L\Omega}^\dagger \,.
\end{equation}
Rather than passively viewing this transformation as an operator redefinition, this can equivalently be viewed as an active transformation made on the state of the link
\begin{equation}
    \ket{\psi} \longmapsto \hat{\Theta}_{L\Omega}^\dagger \ket{\psi} \,.
\end{equation}
Similarly, all link operators $U(n-e_i,e_i)$ entering site $n$ from a negative direction are affected at their right endpoint by a gauge transformation as in \cref{eq:wilson_line_gauge_trans}, which with the help of \cref{eq:U_translation} can be expressed as
\begin{multline}\label{eq:U_right_trans}
    \hat U^j_{mn}(n-e_i,e_i) \longmapsto \\
      \hat{U}^j_{{m}{m'}} D^{j}_{m'n}(\Omega^{-1})
        = \hat\Theta_{R\Omega}^\dagger \hat{U}^j_{mn} \hat\Theta_{R\Omega}.
\end{multline}
Viewed instead as an active transformation on the states of those links, they are transformed via
\begin{equation}
    \ket{\psi} \longmapsto \hat{\Theta}_{R\Omega} \ket{\psi}.
\end{equation}
Note that one should view the transformation either in a passive sense or an active sense, but not both at once. We will choose the active sense in the following.

We can now easily express the operator enacting the gauge transformation at site $n$. 
It is given by
\begin{equation}\label{eq:theta_definition}
    \hat{\Theta}_\Omega(n)
      = \prod_{i=1}^d [\hat{\Theta}_{L\Omega}(n, e_i)^\dagger \hat{\Theta}_{R\Omega}(n-e_i, e_i)] \,.
\end{equation}
As in the case of the individual translation operators, this can always be written as
\begin{equation}
    \hat{\Theta}_\Omega(n) = \exp(i \phi^a(\Omega) \hat{G}^a(n)) \,,
\end{equation}
where $\phi(\Omega)$ are normal coordinates for the group element $\Omega$ and $\hat{G}^a(n)$ is the Hermitian operator generating gauge transformations at site $n$. 
To arrive at an expression for $\hat{G}^a(n)$, note that all operators appearing in the product of \cref{eq:theta_definition} commute with each other, so using \cref{eq:E_def} we find
\begin{equation}\label{eq:Ga_def}
    \hat{G}^a(n) = \sum_{i=1}^d \left[\hat{E}_R^a(n-e_i, e_i) - \hat{E}_L^a(n, e_i) \right] \,.
\end{equation}
This is the lattice analog of the covariant divergence of the chromo-electric field, otherwise the left-hand side of Gauss's law $\mathcal{D}_i E^i = 0$. 
For this reason it is often said that Gauss's law generates gauge transformations.

\begin{figure}[b]
\centering
\includegraphics[width=0.35\textwidth]{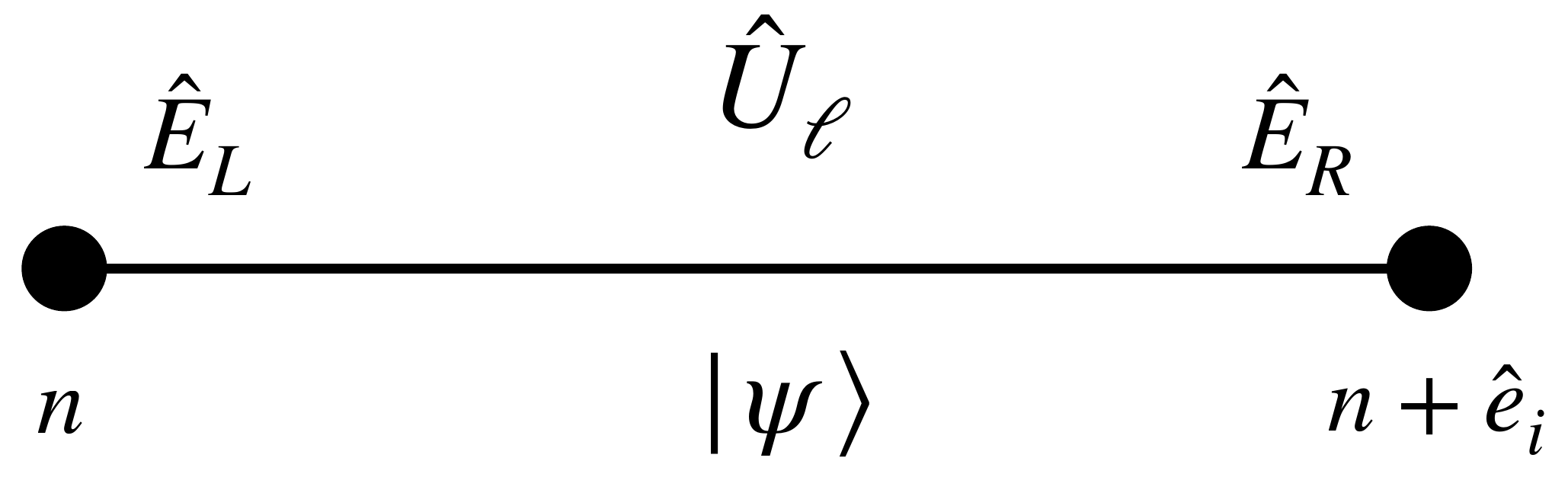}
\caption{The operators at a single link of the Hilbert space, placed according to their gauge transformation properties.}
\label{fig:one_link_hilbert}
\end{figure}

Note however that the lattice version of Gauss's law $\hat{G}^a(n) = 0$ is not true as an operator equation. 
Rather it must be imposed as a constraint on the Hilbert space of states. 
Let $\ket{\Psi}$ be a state of the entire lattice. 
In order that $\ket{\Psi}$ correspond to a physical state, we demand that it satisfies the equation
\begin{equation}\label{eq:gauss_law}
    \hat{G}^{a}(n) \ket{\Psi_\text{phys}} = 0 \,,
\end{equation}
for all sites $n$. This equation is equivalent to the statement that physical states are invariant under all gauge transformations
\begin{equation}
    \hat{\Theta}_\Omega(n) \ket{\Psi_\text{phys}} = \ket{\Psi_\text{phys}} \qc
      \forall\, \Omega(n), \forall\, n \,.
\end{equation}

We now construct the lattice Hamiltonian. 
As previously mentioned, a consequence of Weyl gauge is that the Hamiltonian is gauge invariant.
Switching for the moment back to the passive viewpoint, and working with the operators packaged as $\hat{E}_L = \hat{E}^a_L T^a$ and $\hat{E}_R = \hat{E}_R^a T^a$, the electric operators and link operators transform under gauge transformations via
\begin{align}
    \hat{E}_L(n,e_i) &\longmapsto \Omega(n) \, \hat{E}_L(n,e_i) \, \Omega(n)^\dagger \\[10pt]
    \hat{E}_R(n,e_i) &\longmapsto \Omega(n+e_i) \, \hat{E}_R(n,e_i)  \, \Omega(n+e_i)^\dagger \\[10pt]
      \label{eq:U_gauge_trans}
    \hat{U}(n,e_i) &\longmapsto \Omega(n) \, \hat{U}(n,e_i) \, \Omega(n+e_i)^\dagger \,,
\end{align}
where $\Omega(n)$ is the gauge transformation function (in the defining representation). These gauge transformation properties are visualized in \cref{fig:one_link_hilbert}.

From this we can see that the trace of the quadratic Casimir 
\begin{equation}
    \Tr\hat{E}^2 = \frac{1}{2} \hat{E}_L^a \hat{E}_L^a = \frac{1}{2} \hat E_R^a \hat{E}_R^a
\end{equation} 
is gauge-invariant.

As for the link operators, notice that from \cref{eq:U_gauge_trans} any ordered product of oriented link operators along any closed loop $C$ starting and ending at site $n$ transforms according to
\begin{equation}
    \hat{U}[C] \longmapsto \Omega(n) \hat{U}[C] \Omega(n)^\dagger \,.
\end{equation}
Here $\hat{U}[C]$ is
\begin{equation}
    \hat{U}[C] = \prod_{\ell \in C} (\hat{U}_\ell)^{\sigma_\ell} \,,
\end{equation}
where $\sigma_\ell = 1$ if link $\ell$ is traversed in the positive orientation in $C$ and $\sigma_\ell = -1$ if link $\ell$ is traversed in the negative orientation in $C$.
The smallest non-trivial loop one can traverse on the lattice is a square consisting of four links, called a plaquette, $p$, specified by a site $n$ and two directions $(e_i,e_j)$. 
Then the most local gauge invariant object formed out of link operators is
\begin{align}
\Tr\hat{P}_p &= \Tr \left[\hat{U}(n,e_i) \hat{U}(n+e_i, e_j) \right. \nonumber\\
& \qquad \times \left. \hat U^\dagger (n+e_j, e_i) \hat{U}^\dagger (n,e_j) \right] \,.
\end{align}

The Kogut--Susskind Hamiltonian is the sum of these two terms over every link and plaquette of the lattice, respectively
\begin{equation}\label{eq:H}
    \hat{H} = \hat{H}_E + \hat{H}_B \,,
\end{equation}
where the electric Hamiltonian is:
\begin{equation}\label{eq:HE}
    \hat{H}_E = \frac{g^2}{2a} \sum_\ell \hat{E}^a_\ell \hat{E}^a_\ell \,,
\end{equation}
and the magnetic Hamiltonian is
\begin{equation}\label{eq:HB}
    \hat{H}_B = \frac{1}{2g^2 a} \sum_p \Tr[2I - \hat P_p - \hat P_p^\dagger] \,.
\end{equation}
Here $I = I_{2\times 2}$ is the identity matrix, inserted so that the magnetic Hamiltonian is positive definite. 
The sum over plaquettes is the sum over only positively oriented plaquettes, $i<j$. 
The factors of $g^2$ and $a$ cause this Hamiltonian to approach the continuum Hamiltonian in the limit that the lattice spacing goes to zero $a \to 0$.

\section{Bases and Truncation Schemes}
\label{sec:bases_and_trucation}
Ultimately in order to simulate a lattice gauge theory on a quantum or classical computer one must pick a basis within the Hilbert space with which to represent the operators and the state. 
In addition, for a bosonic variable like a gauge field, the Hilbert space at even a single link is infinite dimensional and must be truncated. 
One needs to picks a basis and truncation scheme such that truncated theory represents the continuous theory well for the observables of interest, and which possesses controlled convergence properties as the number of basis states used increases. 
Other important considerations include the manifestation of Gauss's law in the truncated representation, as well as the loss of certain group properties and commutators (which is generally unavoidable).

Here we briefly review the most common choices and their limitations, before motivating a different basis that we believe is better suited for truncation in the limit of weak bare coupling. 
As will become clearer later, this basis, which we will call the \emph{mixed basis}, can be thought of as a combination of electric and magnetic bases.
Developing a truncation scheme in the mixed basis is the main aim of this paper.

\subsection{Electric Bases}
\label{subsec:electric_bases}
An electric basis is one that diagonalizes the $\hat{E}^2 = \hat{E}_L^2 = \hat{E}_R^2$ operators.
This leaves freedom to specify which additional mutually-commuting operators (\ie, operators that all commute with $\hat{E}^2$) are chosen to be diagonalized.
The most common choice is the $\hat{E}_L^z$ and $\hat{E}_R^z$ operators.
This basis is called the irrep basis in which the states $\ket{j m_L m_R}$ are used as a basis on each link, where $j$ specifies the eigenvalue of the $\hat{E}^2$ operator, while $m_L$ and $m_R$ denote the eigenvalues of the $\hat{E}_L^z$ and $\hat{E}_R^z$, respectively. 

In this basis the electric Hamiltonian is diagonal (with eigenvalues determined by \cref{eq:Eop_electric_basis}), and the link operators behave like raising and lowering operators with matrix elements following the Clebsch--Gordan series
\begin{align}
\label{eq:Uop_elecric_basis}
&\mel{KN_L N_R}{\hat{U}^j_{{m_L}{m_R}}}{JM_L M_R}\\
&\quad = \sqrt{\frac{\dim(J)}{\dim(K)}}\braket{J M_L jm_L }{KN_L} \braket{KN_R}{JM_Rjm_R} \,. \nonumber
\end{align}
This representation is best suited to strong (bare) coupling where the electric Hamiltonian dominates and the lowest lying states are linear combinations of only the lowest lying irreps. 
In this case the standard truncation scheme is to use only states with $j<j_\mathrm{max}$ for some appropriately chosen cutoff $j_\mathrm{max}$. 
This representation also has the advantage that the gauge group is still exactly \sutwo (although the group element basis loses some properties after truncation).
As a result Gauss's law is exactly expressed by $\hat{G}^{a}(n) \ket{\Psi_\text{phys}} = 0$ for the same $\hat{G}^a$, which has an easily interpretable meaning: the state of the $2d$ links emanating from site $n$ must form a singlet. 

Variants of the electric basis exist, such as representations in terms of Schwinger bosons~\cite{Mathur:2004kr,mathur2006loop,mathur2007loop} and the Loop-String-Hadron basis~\cite{Raychowdhury:2018osk,Raychowdhury2023}. 
In these representations the states are not labeled by $\ket{j m_L m_R}$, but they are still eigenstates of the electric operators and truncated in terms of the largest allowed representation.

The main drawback of this basis is that it becomes inefficient at weak coupling, where the magnetic Hamiltonian dominates and the lowest lying energy eigenstates become a sum of a very large number of irreps.

\subsection{Magnetic Bases}
\label{sec:magnetic_basis}

The magnetic basis is the group element basis, in which the eigenstates $\ket{\g}$ of the link operators are the basis states. 
In practice, one picks coordinates $\phi^a(\g)$ for the group and converts all expressions to coordinate expressions. 
For \sutwo, three coordinates are necessary, for instance the three Euler angles $\alpha,\beta,\gamma$ specifying a rotation. One can also work with more than three coordinates with constraints among them (\eg, Cayley--Klein parameters).
The magnetic basis is the starting point for the construction of the mixed basis, for which purpose it will be convenient to now introduce axis-angle coordinates.

\subsubsection*{Axis-Angle Coordinates}

As is familiar from quantum mechanics, elements of \sutwo can be interpreted as rotations in three-dimensional space acting on Pauli spinors. 
Such rotations can be specified by an axis of rotation, given by a unit vector $\hat n$, and a rotation angle, $\omega$. 
Axis-angle coordinates for \sutwo correspond to exactly this parameterization, with a rotation axis $\hat{n} = (\cos\phi \sin\theta, \sin\phi\sin\theta, \cos\theta)$ specified by the spherical coordinates $(\theta,\phi)$, and the rotation about this axis given by $\omega$.

The ranges of the coordinates (so that no point is double-counted) are
\begin{equation}
\begin{split}
    \theta &\in [0,\pi] \\
    \phi &\in [0,2\pi] \\
    \omega &\in [0,2\pi]
    \,.
\end{split}
\end{equation}
Note that the corresponding range for \sothree would be $\omega \in [0,\pi]$, but for \sutwo the value $\omega = 2\pi$ is the single point given by the negative identity matrix.

Explicitly, axis-angle coordinates are defined by the mapping using the defining representation
\begin{multline}
\label{eq:axis_angle}
(\omega,\theta,\phi) \mapsto  D^{1/2}(\omega, \theta, \phi)= e^{-i \omega\, \hat{n}\cdot\bm{\sigma} / 2} \\
= \begin{pmatrix} \cos\frac{\omega}{2} - i\sin\frac{\omega}{2}\cos\theta &-i\sin\frac{\omega}{2} \sin\theta e^{-i\phi} \\[10pt]
-i\sin\frac{\omega}{2} \sin\theta e^{i\phi} &
\cos\frac{\omega}{2} + i\sin\frac{\omega}{2}\cos\theta
\end{pmatrix} \,.
\end{multline}
It is useful to also understand how these coordinates sit on the group manifold, as opposed to simply viewing them as rotations on three-dimensional space. 
As a manifold, \sutwo is diffeomorphic to the three-sphere. This can be seen easily by recalling that \sutwo is the set
\begin{equation}
\mathrm{SU}(2) = \left\{ \begin{pmatrix} a & -b^* \\ b & a^* \end{pmatrix}
\middle\vert \quad a,b\in\mathbb{C}, \ |a|^2 + |b|^2 = 1 \right\}.
\end{equation}
Then the four real numbers $\Re(a), \Im(a), \Re(b), \Im(b)$ are the coordinates for a point on the three-sphere $S^3$ embedded in $\mathbb{R}^4$.

\begin{figure*}[t!]
\centering
\includegraphics[width=0.8\textwidth]{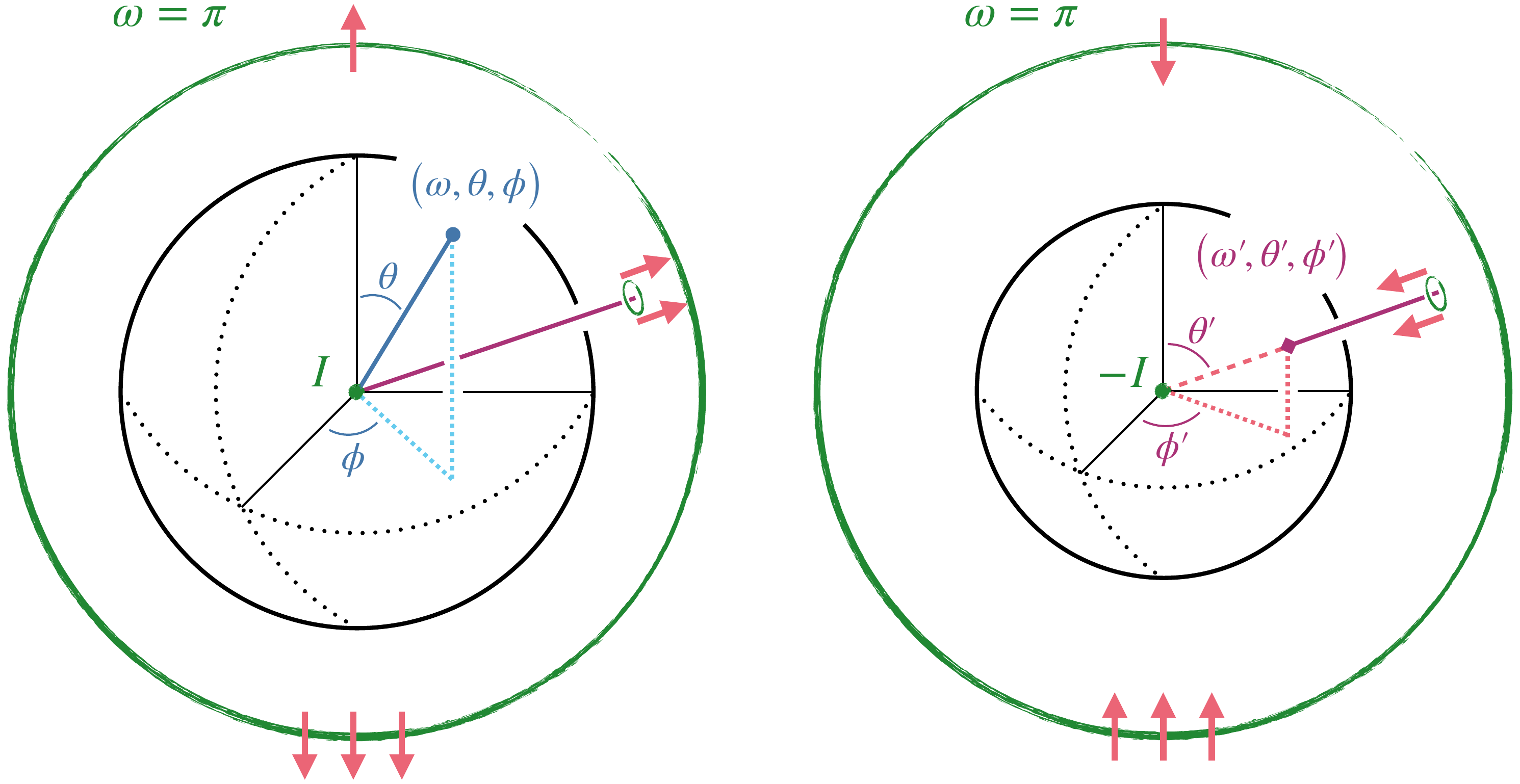}
\caption{Axis-angle coordinates for \sutwo. The north hemisphere (left) and south hemisphere (right) of the manifold are connected by gluing rules (highlighted using reds arrows) at $\omega = \pi$. These glued surfaces are demarcated using green on each hemisphere. The two-sphere is a surface of constant $\omega$. We highlight two specific points. One is the specific point, $(\omega,\theta,\phi)$, with $0 < \omega > \pi$, marked by a blue circle on the north hemisphere. The other is the specific point, $(\omega',\theta',\phi')$, with $\pi < \omega' < 2\pi$, marked by a red square on the south hemisphere.}
\label{fig:axis_angle_coords}
\end{figure*}

Given the mapping $\mathrm{SU}(2) \cong S^3$, axis-angle coordinates are hyperspherical coordinates for $S^3$. 
This can be visualized in the following way. 
Recall that a 2-sphere can be built by gluing together a northern hemisphere and a southern hemisphere, each of which is a disk, \ie, the interior of a 1-sphere (a circle). 
Similarly, the 3-sphere can be built by gluing together a northern hemisphere and a southern hemisphere, each of which is the interior of a 2-sphere, \ie, a solid ball. 
This is visualized in \cref{fig:axis_angle_coords}. 
The north hemisphere is a solid ball (left) with the centermost point (the north pole) corresponding to the identity element of \sutwo.
The south hemisphere is another solid ball (right) with the centermost point (the south pole) corresponding to the negative identity element. 
The two hemispheres are joined by identifying the equatorial surface, i.e. the outer surface of the two balls. The equatorial surface is the sphere of constant $\omega=\pi$ corresponding to all rotations by $\pi$. 
Viewed in this way, the angles $(\theta,\phi)$ are coordinates for 2-spheres of constant radius in \sutwo centered around the identity element, while the angle $\omega$ measures the distance from the north pole.

The Haar measure in axis-angle coordinates is
\begin{align}
    \dd{\g} = 4 \sin^2 \frac{\omega}{2} \sin\theta\, \dd{\omega} \dd{\theta} \dd{\phi} .
\end{align}
Note that the coordinates $(\omega,\theta,\phi)$ differ from the standard convention for hyperspherical coordinates $(\psi,\theta,\phi)$ by the rescaling $\omega/2 = \psi$.
When working in group element coordinates, the electric operators become linear differential operators in the coordinates $\phi^a$. 
Explicit coordinate expressions will be supplied later as needed.

The literature has a number of truncation choices associated with the magnetic basis, but generally all involve some replacement of the continuous group parameter $\g$ with a finite sampling of points $\g_i$. 
As pointed out in~\cite{Lamm2020,Lamm2022,Zohar:2021nyc}, the resulting truncated Hilbert space generally only maintains its group structures if the $\g_i$ form a finite subgroup of \sutwo. 
Ideally one would use a family of finite subgroups of increasing order which converge to the full gauge group, as one can do in the Abelian case where \uone is $\lim_{N\to\infty} \mathbb{Z}_N$. 
However, classification theorems tell us that non-Abelian groups typically have only a finite number of crystal subgroups so such families do not exist (for example \sothree admits only the tetrahedral, octahedral, and icosahedral groups)~\cite{conwaybook2004}.
Nonetheless there has been some work investigating their use~\cite{Alexandru:2019nsa,Lamm2022,Lamm2020,Bender:2018rdp,Alam:2021uuq,Zache:2023cfj,Gonzalez-Cuadra:2022hxt,Ji:2022qvr,Alexandru:2021jpm}.

Alternatively one may simply sample some points $\g_i$, not necessarily forming a group, but concentrated in the support of the wave functions $\Psi(\g)$ one aims to describe.
This is fundamentally the approach taken in this paper. 
If the sampling $\g_i$ is made so as to remain in the support of $\Psi(\g)$ even as the coupling constant $g$ is changed, the accuracy of the representation will be independent of the coupling constant for all values of the coupling constant. 
The main drawback of this approach is the loss of group properties, especially the loss of group composition. 
For this reason, we pair this approach with a gauge-fixing procedure, which is made before truncating the Hilbert space. 
At that point the lack of gauge transformations is irrelevant. 
The price paid for gauge fixing is a non-local electric Hamiltonian, as we will discuss in more detail later.
As will be seen, this non-locality will only lead to interactions between pairs of lattice sites.

In the next section we demonstrate this procedure for the simplest case of a lattice containing a single plaquette, which will then allow us to also describe the choice of basis and the truncation scheme. 
The gauge fixing is made via a maximal tree approach, a construction first developed in~\cite{PhysRevD.15.1128}.
A detailed description of the gauge fixing which generalizes to arbitrary lattice sizes and dimensions is made in \cref{sec:max_tree_gauge_general} with some details supplied in \cref{app:maxTree}.

\section{Lattice With a Single Plaquette}
\label{sec:singlePlaquette}

In this section we study a minimal system consisting of a single plaquette, which we use in order to introduce the gauge fixing procedure and the mixed basis representation without the unnecessary distractions associated with a general lattice. 
The generalization to all lattices is provided afterwards.

\begin{figure}[b]
\centering
\includegraphics[width=0.15\textwidth]{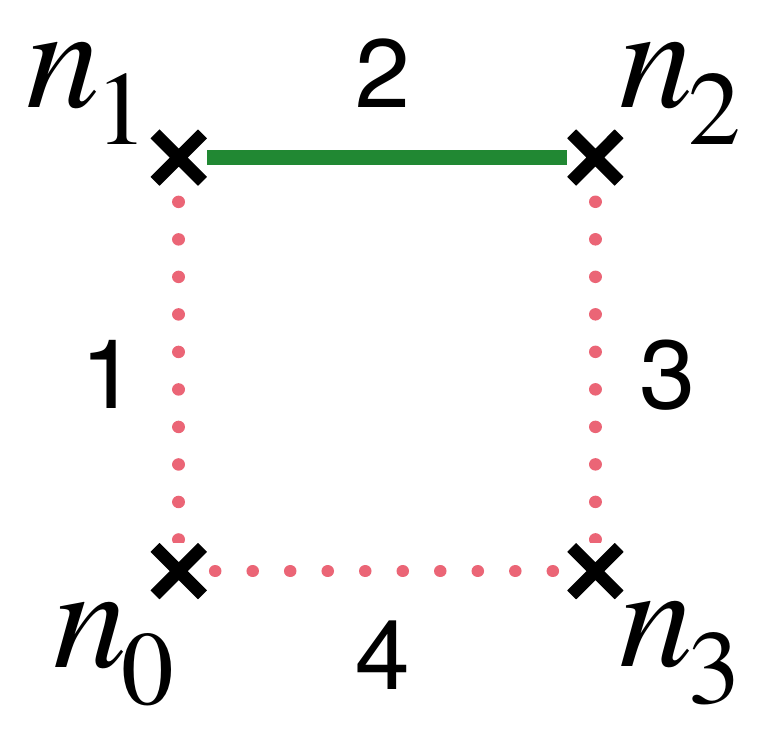}
\caption{A lattice with a single plaquette. The lattice sites are labeled by $n_{0, \dotsc, 3}$. 
Dotted Links in red are part of the maximal tree and will be eliminated. }
\label{fig:single_plaquette}
\end{figure}

A single-plaquette universe with open boundary conditions consists of four lattice sites connected by four links (see \cref{fig:single_plaquette}). 
The Hamiltonian of this system is
\begin{equation}
  \begin{split}
    H_E &= \frac{g^2}{2 a} \sum_{\ell =1}^4 \hat{E}^a_{\ell}\hat{E}^a_{\ell} \\
    H_B &= \frac{1}{2g^2 a} \Tr{2I - \hat{P}_\square - \hat{P}^\dagger_\square}  \,,
  \end{split}
\end{equation}
where the single plaquette operator is $\hat{P}_\square = \hat{U}_1 \hat{U}_2 \hat{U}_3^\dagger \hat{U}_4^\dagger$. 
The links are labelled simply by the index $\ell = 1,2,3,4$, while the sites are labeled by $n_0$--$n_3$.

\subsection{Gauge Fixing}\label{sec:one_plaq_gauge_fix}

Before demonstrating a particular gauge fixing construction, consider for a moment what would be required in order to sample the wave function in a magnetic basis without performing any gauge fixing, especially at weak coupling. 
In the limit that $g$ goes to zero, the magnetic Hamiltonian dominates and energetically favors unfrustrated plaquettes --- that is, plaquettes for which $P_{\square} \approx I$. 
A nominal ground state at zero coupling then takes the form $\Psi(\g_1, \g_2, \g_3, \g_4) = \delta_{\g_1,I} \delta_{\g_2,I} \delta_{\g_3,I} \delta_{\g_4, I}$ with the wave function on each link sharply peaked around zero. 
However, this state is not gauge invariant since a gauge transformation can change any individual $\g_i$ as long as the value of the plaquette is preserved. 
To find a gauge-invariant wave function, the nominal state must be projected into the physical Hilbert space by acting on it with each of the four projectors \cite{Haegeman2015}
\begin{align}
\label{eq:projector}
    \hat{\Pi}(n) = \frac{1}{16 \pi^2} \int\dd{\g} \hat{\Theta}_{\g}(n)
\end{align}
for the four sites of the lattice $n=0,1,2,3$.
Recalling that $\hat \Theta_\g(n)$ acts on the links by group translation, the resulting state will be in a sense smeared over the group.
The primary purpose of gauge fixing is to unsmear the wave function so that it can be sampled efficiently. A more precise description of this smearing will be given in \cref{sec:geometry}.

Of many possible gauge-fixing procedures that would accomplish this goal, we choose to work in a \emph{maximal tree gauge} in which the largest possible number of link variables is eliminated and set equal to the identity (rather than fixing combinations of multiple link variables).
We now carry this out for the one-plaquette universe, with the procedure for arbitrary lattices being given in \cref{sec:max_tree_gauge_general}.

Consider a magnetic basis state $\ket{\g_1 \g_2 \g_3 \g_4}$. 
There are many other basis states that are gauge-equivalent to this one, for instance $\ket{\h \g_1, \g_2, \g_3, \h\g_4}$, which is the result of applying gauge transformation $\Theta_{\h}$ at the origin $n_0$. 
The set of all basis states reachable from $\ket{\g_1 \g_2 \g_3 \g_4}$ by gauge transformations forms an equivalence class of configurations. 
Gauge fixing involves picking a representative sample from each equivalence class with which to represent states. 
The following series of gauge transformations is made in order to arrive at a representative state
\begin{equation}
  \begin{split}
    \ket{\g_1 \g_2 \g_3 \g_4}
      &\xrightarrow{ \hat{\Theta}_{\g_1}(n_1)} \ket{I, \g_1 \g_2, \g_3, \g_4} \\
      &\xrightarrow{ \hat{\Theta}_{\g_4}(n_3)} \ket{I, \g_1 \g_2, \g_4 \g_3 , I} \\
      &\xrightarrow{ \hat{\Theta}_{\g_4 \g_3}(n_2)} \ket{I, \g_1 \g_2 \g_3^{-1} \g_4^{-1}, I, I} \,.
  \end{split}
\end{equation}
In words, one works their way away from the origin, making use of gauge transformations at every site to set as many links as possible to the identity. 
Note that at no point do we make use of the gauge transformation at the origin.

The wave function will now be expressed in terms of a single variable $\g_1 \g_2 \g_3^{-1} \g_4^{-1}$, which is nothing other than the eigenvalue of the plaquette operator $\hat{U}_1 \hat{U}_2 \hat{U}^\dagger_3 \hat{U}^\dagger_4$. 
We rename this operator to be $\hat{X} = \hat{P}_\square$ and now refer to it as the loop operator. 
The remaining gauge transformation at the origin is identified as a global transformation, and affects the loop operator through
\begin{equation}
    \hat{X} \to \hat{\Theta}_{n_0} \hat{X} \hat{\Theta}_{n_0}^\dagger. 
\end{equation}
This equation represents a single remaining global constraint on the wave function, which we will address in a later section.

In terms of the loop operator, the magnetic Hamiltonian is given by
\begin{equation}
\label{eq:gauge_fixed_magnetic_ham}
    H_B = \frac{1}{2g^2 a}\Tr {  2I - \hat{X} - \hat{X}^\dagger } \,.
\end{equation}
Following \cref{app:maxTree}, we define (left) electric operators that have been parallel transported to the origin by
\begin{equation}
  \begin{split}
    \hat{\mathcal{J}}_1 &\equiv \hat{E}_{L1} \\
    \hat{\mathcal{J}}_2 &\equiv \hat{U}_1 \hat{E}_{L2} \hat{U}_1^\dagger \\
    \hat{\mathcal{J}}_3 &\equiv \hat{U}_4 \hat{E}_{L3} \hat{U}_4^\dagger \\
    \hat{\mathcal{J}}_4 &\equiv \hat{E}_{L4}
    \,.
  \end{split}
\end{equation}
The $\hat{\mathcal{J}}$ all have the property that they transform as though located at the origin
\begin{equation}
    \hat{\mathcal{J}}_{\ell}
      \to \hat{\Theta}_{n_0} \hat{\mathcal{J}}_{\ell} \hat{\Theta}_{n_0}^\dagger \,.
\end{equation}
The action of the transported electric operators on the loop variable is
\begin{equation}
\begin{split}
    \comm{\hat{\mathcal{J}}^a_1}{\hat{X}} &= T^a \hat{X} \\
    \comm{\hat{\mathcal{J}}^a_2}{\hat{X}} &= T^a \hat{X} \\
    \comm{\hat{\mathcal{J}}^a_3}{\hat{X}} &= - \hat{X} T^a \\
    \comm{\hat{\mathcal{J}}^a_4}{\hat{X}} &= - \hat{X} T^a \,.
\end{split}
\end{equation}
Being formally the same as the action of left and right electric operators on a link operator, we identify that when acting on physical states we have
\begin{align}
    \hat{\mathcal{E}}_L &= \hat{\mathcal{J}}^a_1 = \hat{\mathcal{J}}^a_2 \, \\
    \hat{\mathcal{E}}_R &= -\hat{\mathcal{J}}^a_3 = -\hat{\mathcal{J}}^a_4 \,,
\end{align}
where $\hat{\mathcal{E}}_L$ and $\hat{\mathcal{E}}_R$ generate left and right transformations on $\hat{X}$. 
Given the above discussion, the electric Hamiltonian becomes
\begin{equation}
    H_E = \frac{2g^2}{a} \hat{\mathcal{E}}^2 \,.
\end{equation}
A state is now represented by a wave function $\Psi(\g)$ over a single group parameter (the eigenvalue of the loop operator $\hat X$), and all local Gauss' law constraints are automatically satisfied. 
The remaining global constraint is
\begin{equation}\label{eq:one_plaq_global_constraint}
    \Psi(\g) = \Psi(\h \g \h^{-1})
\end{equation}
for all $\h \in \mathrm{SU}(2)$. 
The reader may recognize that this constraint is fairly trivial, but for a multi-plaquette system it is generally not as straightforward to satisfy.

\subsection{Representation in the Mixed Basis}
In what follows it will be useful to express the gauge fixed system in axis-angle coordinates (see \cref{sec:magnetic_basis}). 
These coordinates have the property that the conjugacy class of a point in \sutwo is labelled by a single coordinate, $\omega$, which will make it relatively simple to sample the support of the wave function.
In axis-angle coordinates we work with a wave function
\begin{equation}\label{eq:psi_axis_angle}
    \braket{\omega\theta\phi}{\Psi} = \Psi(\omega,\theta,\phi) \,.
\end{equation}

Using \cref{eq:axis_angle,eq:gauge_fixed_magnetic_ham} the magnetic Hamiltonian in axis-angle coordinates is
\begin{equation}\label{eq:HB_single}
    \hat{H}_B = \frac{2}{g^2 a} \pqty{1 - \cos\frac{\omega}{2}} \,.
\end{equation}
The total electric operator $\hat{\mathcal{E}}^2$ is given by the Laplace-–Beltrami operator
\begin{equation}\label{eq:Esq}
    \hat{\mathcal{E}}^2 = \frac{\hat L^2}{4\sin^2\frac{\omega}{2}}
                          - \pdv[2]{\omega} - \cot\frac{\omega}{2} \pdv{\omega} \,,
\end{equation}
with $\hat{L}^2$ denoting the Laplacian on a 2-sphere
\begin{equation}
    \hat{L}^2 = -\pdv[2]{\theta} - \cot\theta\pdv{\theta}
                - \frac{1}{\sin^2\theta}\pdv[2]{\phi} \,.
\end{equation}
The Hamiltonian is then
\begin{equation}\label{eq:single_ham}
  \begin{split}
    H &= \frac{2 g^2}{a} \bqty{\frac{\hat{L}^2}{4\sin^2\frac{\omega}{2}}
 - \pdv[2]{\omega} - \cot\frac{\omega}{2} \pdv{\omega}} \\
& \qquad + \frac{2}{g^2 a} \pqty{1 - \cos \frac{\omega}{2}} \,.
  \end{split}
\end{equation}
The problem is now formally very similar to a problem in three dimensional single particle quantum mechanics, with centrifugal term $\hat{L}^2/4\sin^2\frac{\omega}{2}$ and central potential $V(\omega) = \frac{2}{g^2 a} \pqty{1 - \cos \frac{\omega}{2}}$. 
Once multiple interacting plaquettes are present, the potential ceases to be central.

The wavefunction can always be expanded in spherical harmonics via
\begin{equation}\label{eq:single_wf}
    \Psi(\omega, \theta, \phi)
      = \sum_{\ell, m} c^\ell_m \frac{u^{\ell}_m(\omega)}{2\sin\frac{\omega}{2}}
                         Y^\ell_m(\theta, \phi) \,.
\end{equation}
The factor of $2\sin(\omega/2)$ has been introduced such that with normalized functions $u^\ell_m(\omega)$
\begin{equation}
    \int_0^{2\pi} \dd{\omega} \abs{u^\ell_m(\omega)}^2 = 1 \quad \forall \ell, m \,,
\end{equation}
and normalized coefficients
\begin{equation}
    \sum_{\ell,m} \abs{c^\ell_m}^2 = 1 \,,
\end{equation}
one finds a properly normalized wave function
\begin{equation}
    \int\dd{\g}  \abs{\Psi(\omega, \theta, \phi)}^2 = 1 \,.
\end{equation}
Having defined the radial function $u^\ell_m(\omega)$ in this way, some care in needed to ensure that the wave function does not exhibit unphysical behavior. 
Generally, $\Psi$ will only be differentiable at $\omega=0$ and $\omega=2\pi$ if each $u^\ell_m(\omega)$ has zeros of order at least one at those locations. 
This will become relevant when digitizing this function later.

We can now switch from the magnetic basis $\ket{\omega\theta\phi}$ to the mixed basis $\ket{\omega \ell m}$ via
\begin{equation}
    \bra{\omega' \theta \phi}\ket{\omega \ell m}
      = \frac{\delta(\omega-\omega')}{2 \sin\frac{\omega}{2}} Y^\ell_m(\theta,\phi) \,.
\end{equation}
We refer to this as a \emph{mixed basis} because the states are partially labelled by the magnetic quantum number $\omega$ and partially labelled by the electric quantum numbers $\ell,m$.
These are electric quantum numbers in that they label the eigenvalues of the electric operators $\hat{L}^2$ and $\hat{L}^z$, where $\hat{L}^a \equiv \hat{E}_R^a - \hat{E}_L^a$ (see \cref{app:elec_bases}).
The physical interpretation of these operators will be discussed shortly.

In this basis the wave function is
\begin{equation}
    \braket{\omega \ell m}{\Psi} = c^{\ell}_m u^\ell_m(\omega) \,.
\end{equation}
The representation of the system in this basis is exact before truncation and the main task is simply finding the matrix elements between operators.
For instance the total electric operator is
\begin{multline}\label{eq:Etot_matrix_elements}
\mel{\omega' \ell' m'}{\hat{\mathcal{E}}^2}{\omega \ell m}
= \delta_{\ell\ell'} \delta_{mm'} \delta(\omega-\omega') \\
\times \bqty{-\pdv[2]{\omega} - \frac{1}{4} + \frac{\ell(\ell+1)}{4\sin^2(\omega/2)}} \,.
\end{multline}
Notice that this is fully diagonal in $\ell,m$. 
This will no longer be true in the multi-plaquette problem, where the matrix elements between the individual components of the left and right electric operators are needed. 
They are given in~\cref{sec:multi_plaq_mixed_basis}.
However, those multi-plaquette operators only change the angular numbers by $\pm 1$ or 0, and so the kinetic term is represented by a sparse tridiagonal matrix in the $(\ell,m)$ indices.

\subsection{Solution at Weak Coupling}

In this section we solve the single plaquette problem in the limit $g \to 0$, in which it reduces to a three dimensional simple harmonic oscillator. 
From this we can examine the support of the lowest lying states in radial and angular quantum numbers and understand how to construct a basis with the number of states roughly independent of the coupling constant.

To determine the function $u^\ell_m(\omega)$, one starts from \cref{eq:single_wf} but transforms to the mixed basis as in \cref{eq:Etot_matrix_elements}. %
The Schr\"{o}dinger equation $H \Psi = \lambda \Psi$ then takes the form
\begin{equation}
    \pdv[2]{u^\ell_m}{\omega}
      + \bqty{\frac{1}{4} - \frac{1-\cos\frac{\omega}{2}}{g^4}
              + \frac{\lambda}{2g^2} - \frac{\ell(\ell+1)}{4 \sin^2 \frac{\omega}{2}}}u^\ell_m = 0 \,.
\end{equation}
In the limit $g \to 0$ one can see that the support of the wave function is limited to $\omega \sim g$. 
In order to systematically expand in powers of $g$, one can therefore define the rescaled variable $\hat{\omega}=\omega/g$, giving
\begin{equation}
    \pdv[2]{u^\ell_m}{\hat{\omega}}
      + \bqty{\frac{\lambda}{2} - \frac{\ell(\ell+1)}{\hat{\omega}^2}
              - \frac{\hat{\omega}^2}{8}} u^\ell_m=0 \,,
\end{equation}
where we have dropped subdominant terms in $g$. 
Rearranging slightly, this is the radial equation for a three-dimensional simple harmonic oscillator with `spring constant' $k=1/8$. 
The resulting eigenvalues are
\begin{equation} \label{eq:HO_energies}
    \lambda = \sqrt{2} \pqty{2n + \ell + \frac{3}{2}} \,,
\end{equation}
and the wave functions are simply those of a harmonic oscillator, which up to normalization are
\begin{equation}
    u^\ell_m(\omega) = \omega^{\ell+1} \exp(-\omega^2 / 4\sqrt{2} g^2)
                         L^{(\ell+\frac{1}{2})}_{n}\!\pqty{\frac{\omega^2}{4\sqrt{2}g^2}}
\end{equation}
for associated Laguerre polynomials $L^k_n(x)$.

We see that the angular excitations increase the energy by an amount that does not scale with the coupling constant, and which is of the same order as the radial excitations. 
This indicates that the number of angular states that are needed to represent the low lying spectrum does not scale with $g$. 
The radial function on the other hand is squeezed down into a region of radius $\omMax \approx g$. 
One should then simply adjust the sampling to be concentrated in a region of radius $\min(c g,2\pi)$, where the precise value of $c$ can be chosen to minimize the error introduced by the truncation. 
We will come back to this later.

\subsection{Dealing with the Global Constraint}

As mentioned previously, the maximal tree construction has constrained three of the four possible gauge transformations, but left a global constraint $\Psi(\g) = \Psi(\h\g\h^{-1})$ originating from the gauge transformation at the origin. 
The only way to satisfy this constraint is for $\Psi(\omega,\theta,\phi)$ to only depend on the radial variable. 
To see this, note that the mapping $\g \mapsto \h\g\h^{-1}$ preserves the trace of $\g$, but otherwise sweeps out the conjugacy class of $\g$. 
The conjugacy classes of \sutwo are exactly the 2-spheres parameterized by $(\theta,\phi)$ so the equation reads $\Psi(\omega,\theta,\phi) = \Psi(\omega,\theta',\phi')$.  
Then for the one-plaquette system the physical Hilbert space is just the $\ell=0,m=0$ space. 
The restriction to the physical Hilbert space is not so straightfowardstraightfoward for multi-plaquette systems, and the following discussion is useful for understanding the general case.

The generator of gauge transformations at the origin is $\hat{G}_0^a$, which in the one-plaquette universe is given by
\begin{equation}
    \hat{G}_0^a = - \hat{E}^a_{L1} - \hat{E}^a_{L4} = \hat{\mathcal{E}}^a_R - \hat{\mathcal{E}}^a_L \,.
\end{equation}
We refer to this operator as the total charge of the lattice and denote it by
\begin{equation}
    \hat{L}_\mathrm{tot}^a \equiv \hat{G}_0^a = \hat{\mathcal{E}}^a_R - \hat{\mathcal{E}}^a_L \,.
\end{equation}
The total charge operator inherits the Lie algebra properties of the electric operators
\begin{equation}\label{eq:LL_commutator}
  \comm{\hat{L}_\mathrm{tot}^a}{\hat{L}_\mathrm{tot}^b} = i f^{abc} \hat{L}_\mathrm{tot}^c.
\end{equation}
Being a generator of a global symmetry, it commutes with the Hamiltonian $\comm*{\hat{L}_\mathrm{tot}^a}{\hat{H}} = 0$, as one can easily check. 
Then the situation is identical to a rotationally symmetric system: the simultaneous eigenstates of $\hat{L}_\mathrm{tot}^2$ and $\hat{L}_\mathrm{tot}^z$ label subspaces of the Hilbert space which are not connected by $\hat{H}$. 
These are subspaces of definite total color charge. In the absence of charged matter, only the zero charge sector is physical which reflects the constraint
\begin{equation}
  \hat{L}^a_\mathrm{tot} \ket{\Psi_\text{phys}} = 0 \,.
\end{equation}
Satisfying this constraint for a multi-plaquette system is an $n-$body addition of angular momentum problem, where $n$ is the number of loop operators. 
This will be discussed more later.

After applying the constraint $\hat{L}^a_\mathrm{tot} \ket{\Psi_\text{phys}} = 0$ the Hamiltonian is given purely by a radial equation
\begin{equation}
    H = \frac{2 g^2}{a}\bqty{-\pdv[2]{\omega} - \cot\frac{\omega}{2}\pdv{\omega}}
         + \frac{2}{g^2 a}\pqty{1 - \cos \frac{\omega}{2}} \,,
\end{equation}
which when acting on the rescaled function $u(\omega)\equiv u^0_0(\omega)$ becomes
\begin{multline}
    \bra{\omega'00} H \ket{\omega00} = \delta(\omega-\omega') \\
       \times \bqty{\frac{2g^2}{a}\pqty{-\dv[2]{\omega} - \frac{1}{4}}
                    + \frac{2}{g^2 a}\pqty{1 - \cos \frac{\omega}{2}} }\,.
\end{multline}
In the remainder of this section, we will not enforce the global constraint, since that will help with the generalization to systems with multiple plaquettes.

\subsection{Digitizing the Hilbert Space at All Values of the Coupling}

The discussion of the one-plaquette universe so far has yielded the Hamiltonian
\begin{equation}\label{eq:one_plaq_ham}
    \hat{H} = \frac{2g^2}{a}\hat{\mathcal{E}}^2 + \frac{2}{g^2 a}\pqty{1 - \cos\frac{\omega}{2}} \,,
\end{equation}
which possesses no local gauge transformations, but still has one global gauge transformation.
The single degree of freedom $\hat{X}$ is a Wilson loop beginning and ending at the origin. 
In order to numerically solve this system, one needs to digitize the infinite dimensional Hilbert space and render it finite dimensional.

As discussed in~\cite{macridin:2018,macridin:2018.2,Klco:2019,Bauer:2021gek}, some digitization choices lead to better convergence properties than others, particularly when it comes to representing kinetic operators. 
As a consequence of the Shanon--Nyquist theorem, derivatives represented spectrally using a discrete Fourier transform are exact for functions that are band-limited to within the Nyquist limit. 
On the other hand, functions that have only exponentially small support outside the Nyquist limit will experience only exponentially small error. 
Thus we aim to represent all derivatives using Fourier transforms or their non-Abelian generalizations.

We proceed by digitizing the radial function $u^\ell_{m}(\omega)$ in much the same way as in the \uone case~\cite{Bauer:2021gek}. 
The radial variable is sampled at points
\begin{align}
    \omega_{n} = (n + 1/2) \,\delta \omega , \quad \delta \omega = \omega_{\rm{max}} / N
\end{align}
for $0<n<N-1$. 
The infinite dimensional Hilbert space with states $\ket{\omega}$ labelled by continuous parameter $\omega$ is digitized to include $N$ basis states $\ket{\omega_n}$. 
Rather than sampling over the entire range $0<\omega<2\pi$, we sample only over $0<\omega<\omMax$, where the value of $\omMax$ is appropriately chosen to contain the region of support of $u^\ell_m(\omega)$ and scales with $g$. 
This does not only work at small coupling, but at large coupling as well, as long as one requires $\omMax$ not to exceed $2\pi$. 
In other words
\begin{equation}
\label{eq:omega_max_def}
    \omMax = \min\left( c g, 2\pi\right) \,,
\end{equation}
where the precise value of $c$ can be chosen to minimize the error introduced by the truncation. 
As demonstrated in~\cite{macridin:2018, macridin:2018.2,Bauer:2021gup,Bauer:2021gek}, under the harmonic approximation this optimal value is
\begin{equation}\label{eq:omega_max}
    \omMax =  \min\left( g(N-1)\sqrt{\frac{\sqrt{8}\pi}{N}}, \, 2\pi \right) \,.
\end{equation}
This value is obtained by analytically minimizing $\ev{\comm{\hat{x}}{\hat{p}}-i}$ with respect to the ground state of the harmonic oscillator. 
One could also perform a precalculation to numerically select a value.

The truncated mixed basis for a single loop variable is
\begin{equation}
    \ket{\omega_n \ell m} \qc n=0,\dotsc,N-1, \ell \leq \lMax
    \,,
\end{equation}
which is a space of dimension $N(\lMax+1)^2$. 

One must also select a value for $\lMax$. 
However, this choice is more akin to picking the number of radial basis states $N$.
Generally one would pick $\lMax$ to be as large as can be allowed for the computing architecture, but the computational resources have to be traded between the value of $N$ and $\lMax$. 
We don't expect the value of $\lMax$ to depend on the lattice size.

We now define the truncated operators with respect to the truncated mixed basis. 
The magnetic Hamiltonian is simply
\begin{equation}
    \mel{\omega_{n'} \ell' m'}{\hat{H}_B}{\omega_n \ell m}
      = \delta_{nn'}\delta_{\ell\ell'} \delta_{mm'} h_B(\omega_n) \,,
\end{equation}
where
\begin{equation}
    h_B(\omega_n) = \frac{2}{g^2 a} \pqty{1 - \cos \frac{\omega_n}{2}}
    \,.
\end{equation}
Handling the radial function requires some care. 
Recall that the relationship between the radial functions $u^\ell_m(\omega)$ and the wave function $\Psi$ is
\begin{equation}
\Psi(\omega,\theta,\phi) = \sum_{\ell,m} c^\ell_m \frac{u^\ell_m(\omega)}{2\sin\frac{\omega}{2}}Y^\ell_m(\theta,\phi) \,.
\end{equation}
The factor of $2\sin(\omega/2)$ which serves to flatten the radial integration measure introduces removable singularities in the wave function. 
In order that $\Psi$ remain continuously differentiable at $\omega=0, 2\pi$, the radial function $u^\ell_m(\omega)$ must have zeros of order at least one at $\omega=0, 2\pi$.
Realizing this constraint is equivalent to stating that $u^\ell_m(\omega)$ belongs to the vector space of functions expressible in terms of a pure Fourier Sine series on $[0,2\pi]$.

Given the above, derivatives with respect to $\omega$ are represented via a Fourier transform as
\begin{multline}
\mel{\omega_{n'} \ell' m'}{-\hat{\partial}_\omega^2}{\omega_n \ell m}
 = \delta_{\ell\ell'} \delta_{mm'} \\
\times \sum_{k=0}^{N-1} (\DST^{-1})_{n'k} (p_k^2) (\DST)_{kn}
\,,
\end{multline}
where the unitary discrete Sine transform (DST) is given by
\begin{equation}
(\DST)_{kn} = 
\begin{cases}
\sqrt{\frac{2}{N}} \sin\frac{\pi(n+\frac12)(k+1)}{N} & \text{for $k\neq N-1$} \\
\sqrt{\frac{1}{N}}\sin\pi(n+\frac{1}{2})       & \text{for $k=N-1$}
\end{cases}\,,
\end{equation}
and $p_k$ are the modes present in the Sine transform
\begin{align}
    p_k = \frac{\pi(k+1)}{\omMax} \qc k=0, \dotsc, N-1
    \,.
\end{align}
The total electric operator is then given by
\begin{align}
&\mel{\omega_{n'} \ell' m'}{\hat{\mathcal{E}}^2}{\omega_n \ell m}
 = \delta_{\ell\ell'} \delta_{mm'} \nonumber \\ 
& \qquad\quad \times \bigg[ (\DST^{-1})_{n'k} (p_k)^2 (\DST)_{kn} \\
& \qquad\qquad - \frac{1}{4}\delta_{nn'}
+ \frac{\ell(\ell+1)}{4\sin^2(\omega/2)}\delta_{nn'}\bigg] \, \nonumber
\end{align}
where $k$ is summed. 
Notice that all operators are either diagonal, or formed out of composing Fourier transforms and diagonal operators.

\begin{figure}[h]
\centering
\includegraphics[width=0.47\textwidth]{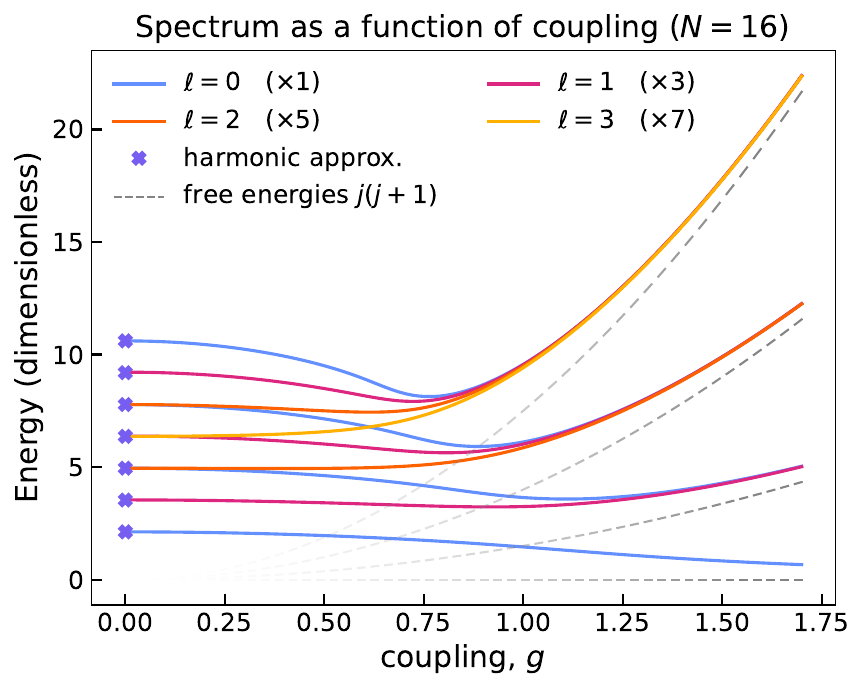}
\caption{The spectrum of the one-plaquette universe as a function of the coupling constant, truncated according to $\lMax = 3$ and $N=16$, which is a $256$ dimensional space requiring 8 qubits. The zero-coupling energies of \cref{eq:HO_energies} are plotted as $\texttt{x}$'s. The strong coupling energies are $2g^2 j(j+1)$, for $j=0,\frac{1}{2},1,\dotsc$ and degeneracy $(2j+1)^2$ (see \cref{eq:Eop_electric_basis}). They are plotted as dotted lines. Note that a line labelled by $\ell$ has degeneracy $(2\ell+1)$, indicated in the legend by, \eg, $(\times 3)$ for a threefold degenerate state. }
\label{fig:one_plaquette_spectrum}
\end{figure}

The spectrum of the truncated system at all values of the coupling can be found by numerically diagonalizing the Hamiltonian, as shown in \cref{fig:one_plaquette_spectrum} for the one-plaquette universe using a modest $N=16$ field values, corresponding to using 4 qubits for the radial Hilbert space. 
At weak coupling one recovers the simple harmonic oscillator states, while at strong coupling the states approach the electric eigenstates $\ket{j \ell m}$\footnote{For a fixed value of $j$, the states $\ket{j{m_L}{m_R}}$ and $\ket{j\ell m}$ span the same space of electric eigenstates. However, they diagonalize different sets of electric operators: as mentioned in \cref{sec:basics_KG}, $\ket{j{m_L}{m_R}}$ diagonalizes the operators $\hat{E}^2$, $\hat{E}^z_L$, and $\hat{E}^z_R$, while $\ket{j \ell m}$ diagonalizes $\hat{E}^2$, $\hat{L}^2$, and $\hat{L}^z$. See \cref{app:elec_bases} }. 

\begin{figure}[h]
\centering
\includegraphics[width=0.5\textwidth]{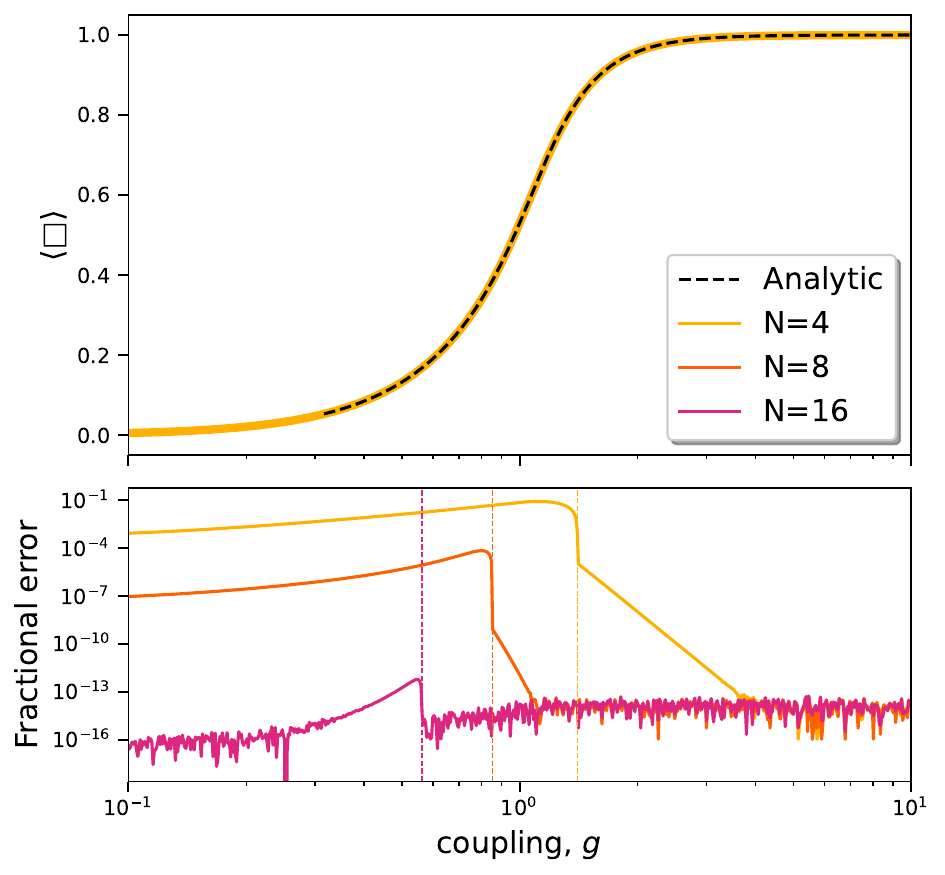}
\caption{(Top) The expectation value $\ev{\square}_\text{exact}$ with respect to the ground state of the total Hamtiltonian, for various values of the coupling. The analytic solution is obtained in terms of a certain Mathieu function (see \cref{app:analytic_solution}). Due to the difficulty of evaluating Mathieu functions at large $q$ values, the exact solution is not plotted at all values. (Bottom) The fractional error $(\ev{\square}_\text{exact} - \ev{\square})/\ev{\square}_\text{exact}$ in evaluating this quantity using a truncation with $N$ radial samples. The vertical lines mark the values of $g$ where the $\min()$ function in \cref{eq:omega_max} is saturated. }
\label{fig:one_plaquette_HB}
\end{figure}

Another useful numerical benchmark is the normalized expectation value of the magnetic Hamiltonian with respect to the ground state of the full system:
\begin{align}
    \ev{\square} = \frac{g^2 a}{2} \mel{\Psi_0}{H_B}{\Psi_0}.
\end{align}
In the \uone case this was considered in \cite{Muschik2021,Haase:2020kaj,Bauer:2021gek} and served as a useful point of comparison. 
The result for this work is shown in \cref{fig:one_plaquette_HB} for various truncation values, along with an analytic solution which is available for the one-plaquette universe (See \cref{app:analytic_solution}). 
Even with only $N=4$ samples, corresponding to 2 qubits, the truncation scheme results in per-mill level accuracy at all values of the coupling. 
As expected on the grounds of the Shanon--Nyquist theorem, after a certain point convergence is exponential in the number of samples used. 
As a result, errors associated with digitization and truncation are not expected to make a significant contribution to the overall error when performing simulations of lattice gauge theories.
%

\section{Lattice with Multiple Plaquettes}
\label{sec:multiPlaquette}

In this section we generalize the preceding discussion to a $d$-dimensional hypercubic lattice containing $N^d$ lattice sites and consider open boundary conditions. 

After gauge fixing, we will arrive at an expression for the Hamiltonian in terms of the dynamics of a complete set of loop operators. 
The matrix elements of the operators appearing in the Hamiltonian are given with respect to the mixed basis. 
As in the case of the one-plaquette universe, we will show how to define a truncated representation of the Hamiltonian with resource requirements that do not scale with the coupling constant. 

All operators appearing in the Hamiltonian will be direct products and compositions of Fourier transforms, diagonal matrices, and tridiagonal matrices.

\subsection{Gauge Fixing}
We begin with the Kogut--Susskind Hamiltonian given in \cref{eq:H,eq:HE,eq:HB}
\begin{equation}
    \hat{H} = \frac{g^2}{2a} \sum_\ell \hat{E}^a_\ell \hat{E}^a_\ell
            + \frac{1}{2g^2 a} \sum_p \Tr[2I - \hat{P}_p - \hat{P}_p^\dagger] \,,
\end{equation}
where the sum over $\ell$ is a sum over all links $\ell(n,e_i)$ in the hypercubic lattice, and the sum over $p$ is a sum over all positively oriented plaquettes $p(n,e_i,e_j)$.  

This Hamiltonian is accompanied by the supplementary Gauss' law constraints
\begin{equation}
    \hat{G}^a(n) \ket{\Psi_\text{phys}} = 0 \qc \forall n
\end{equation}
for each lattice site $n$. 

In this section we will carry out the maximal tree gauge fixing for a general lattice, with some details relegated to \cref{app:maxTree}. 
Before doing so, we will take a moment to provide a geometrical description of gauge fixing and the smearing of the wave function alluded to in \cref{sec:one_plaq_gauge_fix}.

\subsubsection{Geometrical Picture of Gauge Fixing}\label{sec:geometry}

For this purpose it is natural to use the language of fiber bundles to describe the geometrical structure imposed on configuration space by the action of gauge transformations. 
No facility with the mathematical machinery is necessary to appreciate the basic construction.

\begin{figure}[t]
\centering
\includegraphics[width=0.45\textwidth]{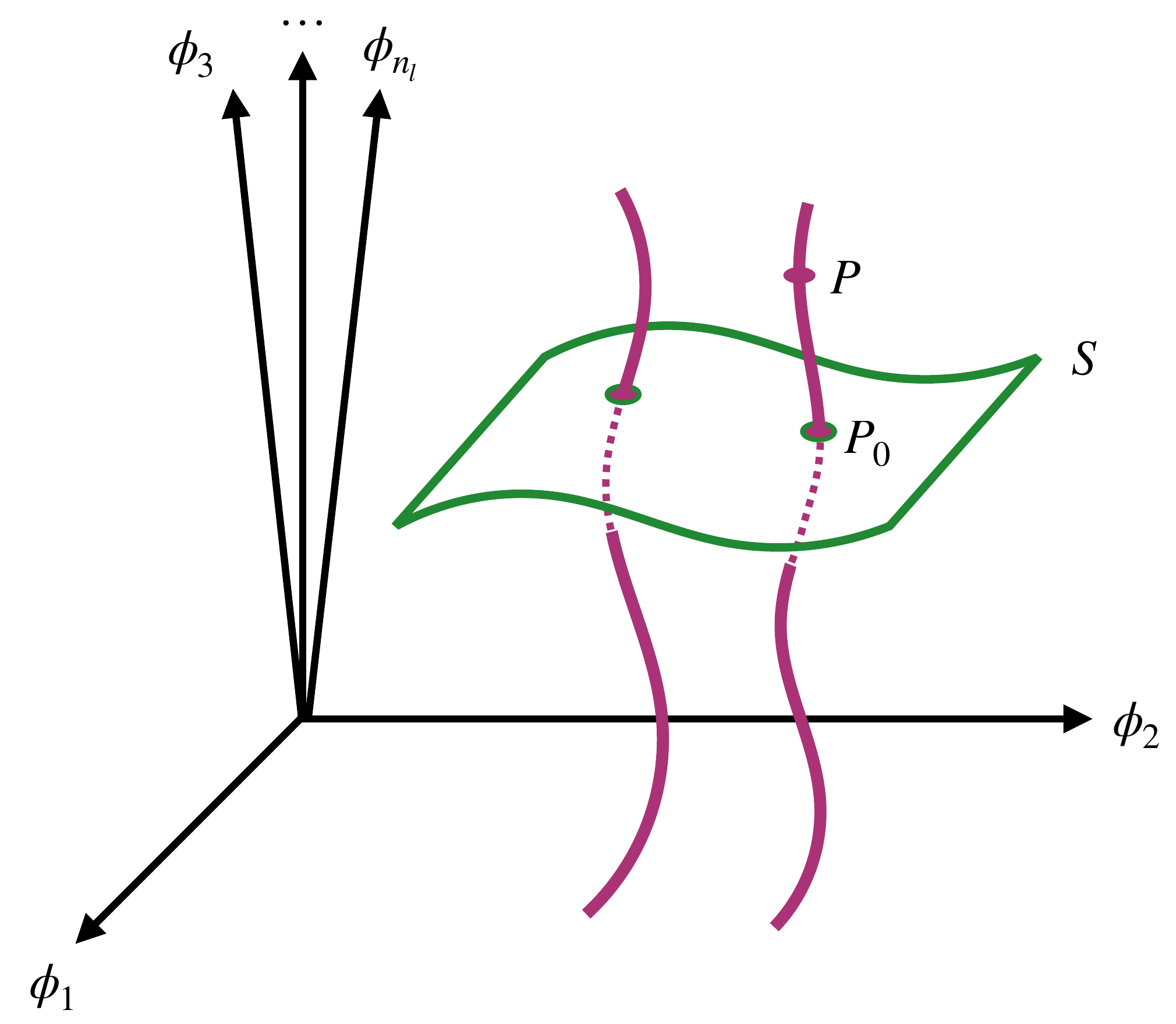}
\caption{The $3n_\ell$ dimensional configuration space of the Kogut--Susskind Hamiltonian is represented using several axes. A typical configuration (point $P$) is acted on by gauge transformations, producing the orbit shown schematically as a line passing through $P$. The $3(n_\ell - N^d + 1)$ dimensional surface $S$ of representative points establishes a gauge convention. This figure is a reproduction of a figure found in~\cite{RevModPhys.69.213}.}
\label{fig:fibers}
\end{figure}

Let the number of links on the lattice be $n_\ell$. 
We begin by considering the $3n_\ell$-dimensional configuration space of the Kogut--Susskind Hamiltonian, illustrated schematically in \cref{fig:fibers}, upon which the parameters $\phi^a_i = \phi^a(\g_i)$ are coordinates. 
Let $P$ be some specific configuration (point in configuration space), and consider the set of configurations that are generated by letting all gauge transformations at one site $n$ act on $P$. 
These configurations form a surface in configuration space passing through $P$. This surface is represented schematically as a line in \cref{fig:fibers}, but its dimension is in fact three. 
This surface is the orbit of the point $P$ under the action of gauge transformations at site $n$. 
The orbit is a copy of the gauge group \sutwo, since the action of a group on itself by left or right translation is free.

The configuration space $Q$ is divided up or foliated into a $3n_\ell-3$ dimensional family of three dimensional orbits, each of which is a copy of \sutwo. 
The orbit structure imposed on configuration space by gauge transformations at site $n$ qualifies this space as a principal fiber bundle over \sutwo. 
In this case \sutwo is called the \emph{structure group}, the quotient space $Q/\text{SU}(2)$ is the \emph{base space}, and the orbits are the \emph{fibers}.

Consider now the orbit of the configuration $P$ under the full gauge group
--- that is, under all gauge transformations at all sites except those at the origin. 
The reason we do not include gauge transformations at the origin is that they will be re-interpreted as a global symmetry related to global charge conservation. 
This will become more clear when performing the maximal tree gauge fixing.
Regardless, the orbit of $P$ under this full gauge group is simply the direct product of \sutwo with itself for every lattice site (except the origin). 
This is because gauge transformations made at different sites commute with each other. 
Then in this case the orbit is a surface of dimension $3(N^d-1)$ passing through point $P$, also called the gauge fiber on which $P$ lives.
Each point on a fiber is gauge equivalent to each other point on that fiber, and they represent the same classical physical configuration.

The fiber bundle structure provides a geometrical picture for the meaning of a quantum state being physical. 
Let $\Psi(\{\phi^a(\g_i)\}) = \bra{\g_0,\dotsc,\g_{n_\ell-1}}\ket{\Psi}$ be the wave function viewed as a function over configuration space. 
Then the Gauss's law constraint
\begin{align}
    \hat{\Theta}_\Omega(n) \ket{\Psi_\text{phys}} = \ket{\Psi_\text{phys}} \qc
    \forall \Omega(n), \forall n \neq n_0
\end{align}
states that a physical wave function is simply one which is constant on each fiber.

We similarly get a geometrical picture for the meaning of gauge fixing. 
To see this, consider the fiber passing through point $P$, and declare some point $P_0$ to be representative of the fiber as a whole. Having done this for one fiber, we extend the choice of a representative point to neighboring fibers, continuing to move outwards in a continuous manner until all fibers have been covered. 
In this way, a $3(n_\ell - N^d + 1)$ dimensional surface $S$ of representative points is swept out. 
The surface $S$ is called a \emph{section} of the fiber bundle, and it is the geometrical interpretation of a gauge fixing procedure. 
Picking a different gauge-fixing convention involves picking a different section $S'$. Generally the only restriction on a section is that it is nowhere tangent to the fibers.

Because physical wave functions are constant on the fibers anyways, we can work with a wave function $\Psi$ over only the gauge section $S$ and extend it uniquely to a physical wave function by making it constant on every fiber. 
This also simplifies the matter of sampling the wave function, as all sampling done along the direction a fiber would be redundant.

\subsubsection{Maximal Tree Gauge}\label{sec:max_tree_gauge_general}

We now pick a particular gauge section and express the Hamiltonian using operators that act only on wave functions defined on that section. 
In the process we will produce a formulation of the theory in terms of a collection of interacting Wilson loops with non-local interactions among them. 
Such non-local interactions are a generic feature of gauge-fixed formulations of gauge theories. 
The particular choice of gauge-fixing we employ belongs to a class of gauges collectively called maximal tree gauges, which were introduced in~\cite{PhysRevD.15.1128}. 
The construction here mostly follows that of~\cite{Ligterink_2000}, which we illustrate mostly for the two-dimensional case, but the generalization to higher dimensions is clear.

\begin{figure}[t]
\centering
\includegraphics[width=0.4\textwidth]{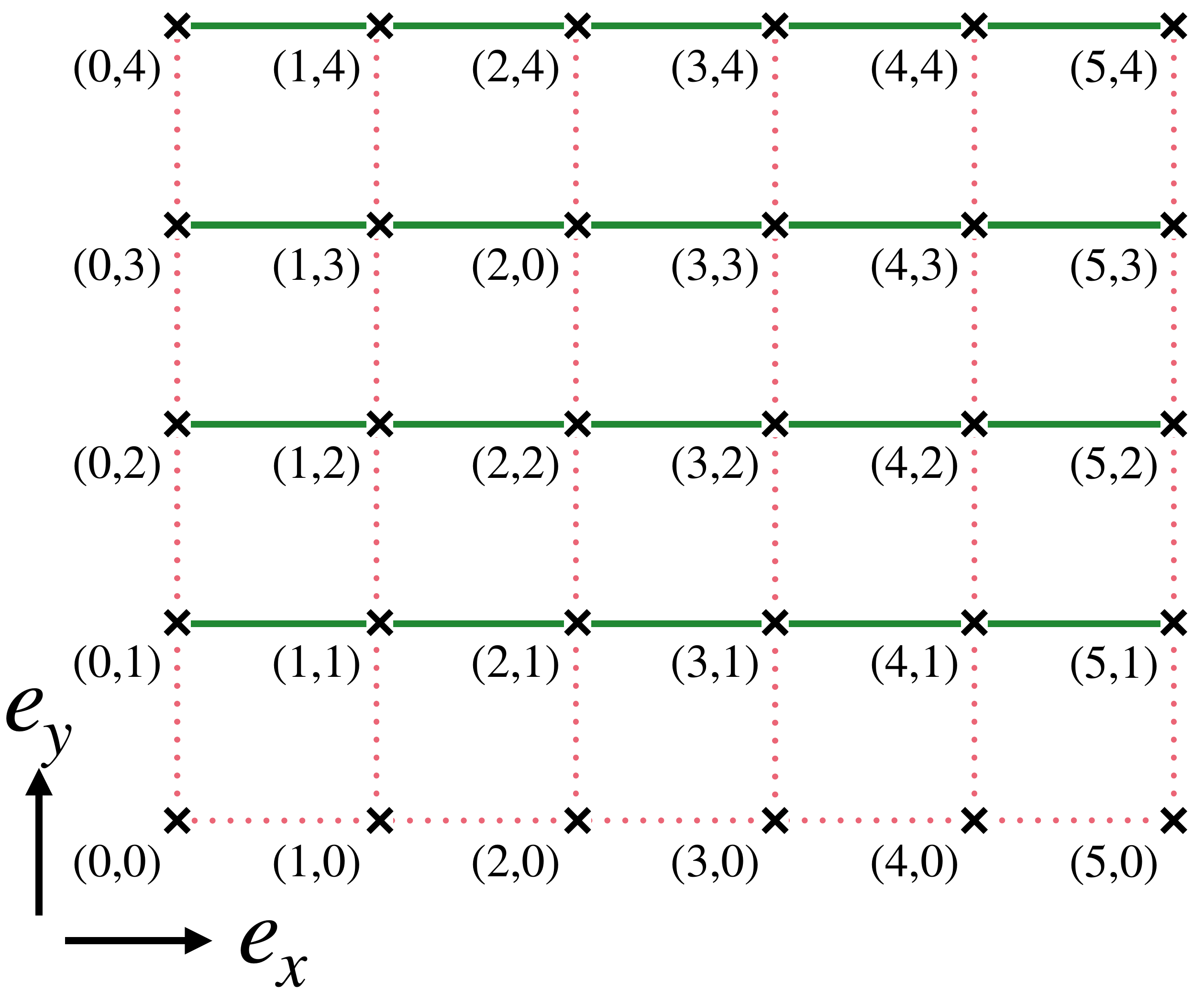}
\caption{A two-dimensional lattice, with a maximal tree identified by the red dotted links, with links not on the maximal tree shown in solid green}
\label{fig:2dLattice}
\end{figure}

To start, we partition the links of the lattice into two disjoint sets referred to as `physical links' and `tree links'. 
The links making up the tree form a tree in the graph sense --- there are no closed loops that can be traversed via tree links. 
In addition, the tree links form a maximal tree, meaning the addition of any link to the tree would create a closed loop. 
All links that are not part of the maximal tree are declared physical links, which we denote by $\kappa$, as opposed to $\ell$ which denotes any link.

There are many different choices possible for a maximal tree, and this work chooses the maximal tree in 2-dimensions as links in the $x$ directions for $y=0$, as well as all links in the $y$ direction. 
The maximal tree used is illustrated in \cref{fig:2dLattice} by the collection of red links. 
In a three-dimensional lattice one would choose the links in the $x$ direction for $y = z = 0$, the links in the $y$ direction for $z = 0$ and all links in the $z$ direction. 
This is the maximal tree definition used in~\cite{Kaplan:2018vnj,MULLER198449, PhysRevD.31.2020, Ligterink_2000}.

We now specify a gauge section using the maximal tree. 
Consider a point $P$ in configuration space along with all the gauge-equivalent points that make up its fiber. 
To establish the representative element on this fiber, we make a series of gauge transformations to set the link variable on every link on the maximal tree equal to the identity element. 
For instance one uses the gauge transformation at site $(0,1)$ to rotate the link variable living on the link connecting $(0,0)$ and $(0,1)$ equal to the identity. 
Next one uses the transformation at site $(0,2)$ to set the link variable living on the link connecting $(0,1)$ and $(0,2)$ equal to the identity. 
This process continues, using gauge transformations at sites moving outward from the origin, until we arrive at the unique point $P_0$ on the fiber where every link variable living on the tree is equal to the identity. 
This configuration is declared representative of the fiber. 
Next this process is repeated for every fiber, which defines the section $S$ specifying the gauge-fixing.

\begin{figure}[b]
\centering
\includegraphics[width=0.4\textwidth]{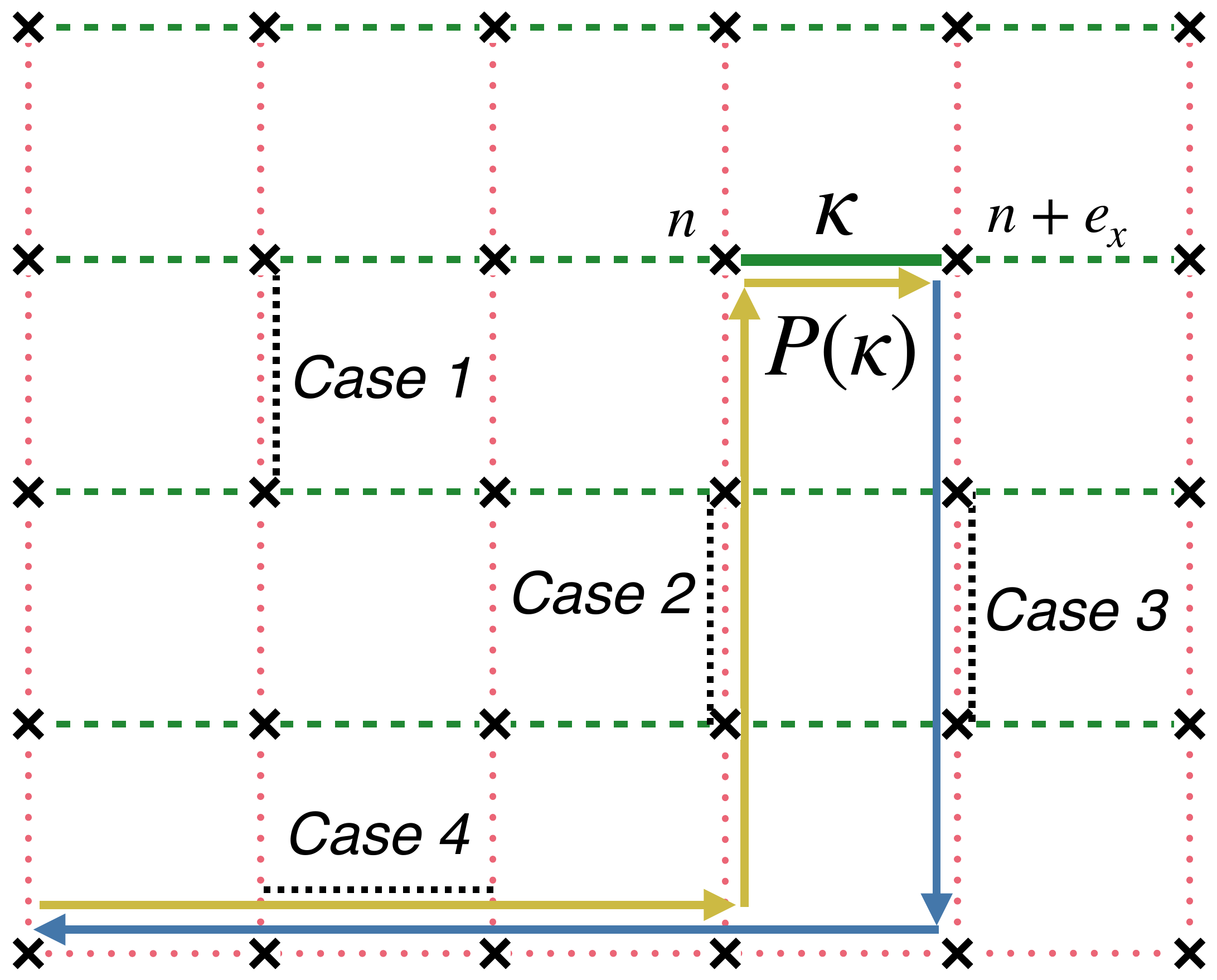}
\caption{The path $P(\kappa)$ for a physical link $\kappa$. All paths start and end at the origin of the lattice $n_0 = (0,0)$. Also indicated are the four possible cases of links when considering the commutation relations between electric operators and the loop operator. Note that links that are removed via the maximal tree convention are shown in dotted red lines while all physical links, except for physical link $\kappa$, are shown with dashed green lines.}
\label{fig:loop1}
\end{figure}

Let us express the representative point $P_0$ in the group-element basis. 
A starting point on the fiber is $P=\ket{\{\g_\tau\},\{\g_\kappa\}}$ where the $\{\g_\tau\}$ belong to the maximal tree and the $\{\g_\kappa\}$ belong to the physical links. 
Sliding along the fiber from $P$ to $P_0$ results in the state $\ket{\{I\},\{X_\kappa\}}$ which has all tree elements equal to the identity, and all physical variables set to a certain product of the original $\g_\ell$ and $\g_\kappa$ variables. 
By following along the series of gauge transformations, one can see that $X_\kappa$ is the oriented product of $\g$'s around the closed loop starting and ending at the origin which traverses the physical link $\kappa$ in the positive orientation and otherwise only traverses the tree. 
This is a unique loop because there is a unique way to get between any two points on the lattice via the maximal tree if we demand each link is only traversed once. 
Such a loop is denoted $P(\kappa)$, and an example loop is illustrated in \cref{fig:loop1}.  

In order to work with operators that act only on the maximal tree section $S$, we then define loop operators
\begin{equation}
    \hat{X}(\kappa) = \prod_{\ell \in P(\kappa) } \hat{U}(\ell)^{\sigma_\ell} \,,
\end{equation}
where $\sigma_\ell$ is $+1$ if the link $\ell$ is traversed in the positive orientation, and $-1$ if it is traversed in the negative orientation. 
Since $\hat X(\kappa)$ is a product of link operators that start and end at the origin, it depends only on gauge transformations made at the origin
\begin{align}
\label{eq:global_gauge_y}
    \hat X(\kappa) \to \Omega(n_0) \hat X(\kappa) \Omega(n_0)^\dagger
    \,.
\end{align}
Clearly $\Tr\hat X(\kappa)$ is gauge invariant.

We can now express the magnetic Hamiltonian in terms of the loop operators. 
Note that a feature in two dimensions is that the number of physical links is equal to the number of plaquettes, while in higher dimensions there are more plaquettes than loop operators. 
In any case, each plaquette is comprised of four links, at least one of which has to be a physical link. 
For a two-dimensional lattice the maximum number of physical links for each plaquette is two, while in higher dimension one can have up to four physical links per plaquette.
In the example given in \cref{fig:2dLattice} the plaquettes in the bottom row contain a single physical link, while all others contain two physical links. 
Given the set of physical links, the plaquette operator can be written in terms of the loop operators as
\begin{align}
\label{eq:P_through_X_app}
    \Tr \hat P_p \equiv \Tr \prod_{\kappa\in p} \hat X(\kappa)^{\sigma(\kappa)}
    \,,
\end{align}
where the product runs over the physical links of the plaquette, the operators $\hat X(\kappa)$ are the loop operators corresponding to each physical link, and $\sigma(\kappa)$ is $+1$ $(-1)$ if the link $\kappa$ is traversed by the plaquette in the positive (negative) direction. 
Note that only the trace of the two sides are equal, since the $\hat X$ operators are parallel transported to the origin.

The magnetic Hamiltonian can easily be obtained using \cref{eq:P_through_X_app}, and one finds in terms of the operators $\hat X$
\begin{align}
\label{eq:HB_max_tree}
    H_B = \frac{1}{2g^2 a}\sum_p \Tr \left( I - \prod_{\kappa\in p} \hat X(\kappa)^{\sigma(\kappa)} \right) + \mathrm{h.c.}
    \,,
\end{align}
which as discussed involves at most two loop operators $\hat X(\kappa)$ for each plaquette $p$ in two dimensions, and at most four loop operators in higher dimensions. 
In this sense the magnetic Hamiltonian is still quasi-local.

The electric Hamiltonian can similarly be expressed in terms of electric operators conjugate to the loop operators. 
For each loop operator $\hat X(\kappa)$, let $\mathcal{E}_L^a(\kappa)$ and $\mathcal{E}_R^a(\kappa)$ be the operators enacting left and right transformations on $\hat X(\kappa)$
\begin{align}
    [\hat{\mathcal{E}}_L^a(\kappa), \hat X(\kappa')] &= T^a \hat X(\kappa) \delta_{\kappa,\kappa'} \,,\\[10pt]
    [ \hat{\mathcal{E}}_R^a(\kappa), \hat X(\kappa')] &= \hat X(\kappa) T^a \delta_{\kappa,\kappa'}
    \,.
\end{align}
A detailed explanation of the relationship between these and the original electric operators is given in \cref{app:maxTree}. 
Here we simply quote the result for the electric Hamiltonian. 
Let $t_{+}(\ell)$ be the set of all physical links $\kappa$ such that $\ell$ is contained in path $P(\kappa)$ as a positive link. 
Similarly let $t_{-}(\ell)$ be the set of all physical links $\kappa$ such that $\ell$ is contained in path $P(\kappa)$ as a negative link. 
Then the electric Hamiltonian is:
\begin{align}
\label{eq:HE_max_tree}
    H_E = \frac{g^2}{2 a} \sum_{\ell} \left( \sum_{\substack{\kappa \in t_+(\ell)}} \hat{\mathcal{E}}^a_{L\kappa} - \sum_{ \substack{ \kappa \in t_{-}(\ell)} } \hat{\mathcal{E}}^{a}_{R\kappa} \right)^2
    \,.
\end{align}
The electric Hamiltonian is still bilinear in electric operators, but now has terms which couple all loop operators, in general. %
As a result the electric Hamiltonian is now non-local.

The total Hamiltonian is given by the sum of \cref{eq:HB_max_tree,eq:HE_max_tree}. 
It contains no local gauge transformations, but has a global transformation generated by the residual gauge transformation at the origin
\begin{align}
\label{eq:gauge_trafo_at_origin}
\begin{split}
    \hat X(\kappa) \to \Omega(n_0) \hat X(\kappa) \Omega(n_0)^\dagger, \\
    \hat{\mathcal{E}}_L(\kappa) \to \Omega(n_0) \hat{\mathcal{E}}_L(\kappa) \Omega(n_0)^\dagger, \\
    \hat{\mathcal{E}}_R(\kappa) \to \Omega(n_0) \hat{\mathcal{E}}_R(\kappa) \Omega(n_0)^\dagger
    \,.
\end{split}
\end{align}

\subsection{Representation in the Mixed Basis}
\label{sec:multi_plaq_mixed_basis}
We now carry out the program of expressing this gauge-fixed Hamiltonian in terms of its matrix elements with respect to the mixed basis. 
This leads to an exact representation of the system. 
We will then need to truncate this basis to arrive at a finite representation amenable to simulation. 
The process is largely the same as in the one-plaquette universe, with the only significant difference coming from the appearance of operators that change the angular quantum numbers $\ell,m$.

The Hilbert space of the Hamiltonian \cref{eq:HB_max_tree,eq:HE_max_tree} is the direct product of a one-link Hilbert space $\otimes_\kappa L^2(G,d\g)$ for each loop operator $\hat X(\kappa)$. 
On each copy of the one-link Hilbert space we employ the mixed basis $\ket{\omega \ell m}_{\kappa}$, and the direct product of all such basis vectors forms a basis for the entire Hilbert space. 
A sample basis vector for the entire Hilbert space is
\begin{align}
\ket{\omega,\ell,m}_{1} \otimes \hdots \otimes \ket{\omega,\ell,m}_{n_\kappa}
\,,
\end{align}
where $n_\kappa$ is the total number of loop operators.

The matrix elements of the operators appearing in the Hamiltonian can be found in a relatively straightforward manner by first expressing them in axis-angle coordinates, and then transforming to the mixed basis.

We begin with the electric operators. 
In coordinates, the electric operators acting on loop $\kappa$ are differential operators with respect to axis-angle coordinates $(\omega,\theta,\phi)_\kappa$. 
From \cref{app:maxTree}, these are 

\begin{align}
\label{eq:ELER_rel}
    \mathcal{E}^a_L = \frac{1}{2}(\Sigma^a - L^a), \quad \mathcal{E}^a_R = \frac{1}{2}(\Sigma^a + L^a)
    \,,
\end{align}
where $L^a$ denote the well known angular momentum operators 

\begin{align}
\label{eq:Lop}
    \hat L^x &= i \left( \sin\phi \frac{\partial}{d \theta} + \cot\theta \cos\phi \frac{\partial}{{\partial} \phi}\right)
    \nonumber\\
    \hat L^y &= i \left( -\cos\phi \frac{\partial}{{\partial} \theta} + \cot\theta \sin\phi \frac{\partial}{{\partial} \phi}\right)
    \nonumber\\
    \hat L^z &= -i \frac{\partial}{{\partial} \phi}
    \,,
\end{align}
and  

\begin{align}
\label{eq:Sigmaop}
    \hat \Sigma^x &= 2i \sin\theta \cos\phi \frac{\partial}{{\partial} \omega} + i \cot\frac{\omega}{2} \nonumber\\
    & \qquad \times \left(\cos\theta \cos\phi \frac{\partial}{{\partial} \theta} - \csc\theta \sin\phi \frac{\partial}{{\partial} \phi}\right)
    \nonumber\\
    \hat \Sigma^y &= 2i \sin\theta \sin\phi \frac{\partial}{{\partial} \omega} + i \cot\frac{\omega}{2} \nonumber\\
    & \qquad  \times\left(\cos\theta \sin\phi \frac{\partial}{{\partial} \theta} + \csc\theta \cos\phi \frac{\partial}{{\partial} \phi}\right)
    \nonumber\\
    \hat \Sigma^z &= 2i \cos\theta \frac{\partial}{{\partial} \omega}- i \cot\frac{\omega}{2}\sin\theta \frac{\partial}{{\partial} \theta}
    \,.
\end{align}
It will be slightly more convenient to work with the raising and lowering operators

\begin{align}
    \hat{\mathcal{E}}^0 &= \hat{\mathcal{E}}^z \\
    \hat{\mathcal{E}} ^{\pm} &= \mp \frac{1}{\sqrt{2}} \left( \hat{\mathcal{E}}^x  \pm i \hat{\mathcal{E}}^y  \right)
    \,,
\end{align}
also called the (covariant) spherical components. 

We now simply let these operators act on the mixed basis states $u^\ell_m(\omega) / (2 \sin\omega/2) Y^\ell_m(\theta,\phi)$ and use properties of the spherical harmonics to re-express the result as a linear combination of spherical harmonics.

Note that all electric operators are vector operators under rotations generated by $\hat L$, so the Wigner-Eckart theorem applies. 
The matrix elements of all vector operators can be expressed as

\begin{align}
\label{eq:vec_ops}
    \bra{\omega'\ell'm'}\hat V^q \ket{\omega \ell m} = \langle \omega'\ell' \vert \vert \bm{V} \vert\vert \omega \ell \rangle \bra{\ell' m'}\ket{\ell 1 m q}
    \,,
\end{align}
where $q=+1,0,-1$ are the three spherical components, $\bra{\ell' m'}\ket{j 1 m q}$ is a Clebsch–Gordan coefficient, and $\langle \omega'\ell' \vert \vert \bm{V} \vert\vert \omega \ell \rangle$, referred to as a reduced matrix element,  has no dependence on the $m',m, q$ quantum numbers.
For the electric operators, the reduced matrix elements are given by

\begin{align}\label{eq:elec_mat_elements}
\begin{split}
    & \langle \omega'\ell' \vert \vert \hat{\bm{\mathcal{E}}}_{R(L)} \vert\vert \omega \ell \rangle = \delta(\omega'-\omega)\\
 & \qquad \times
    \begin{cases}
    -i \sqrt{\frac{\ell}{2\ell'+1}} \big( \frac{\df }{\df \omega} + \frac{\ell}{2}\cot\frac{\omega}{2} \big) & \ell'=\ell-1 \\[15pt]
    +(-)\frac{\sqrt{\ell(\ell+1)}}{2} & \ell'=\ell \\[15pt]
    -i \sqrt{\frac{\ell'}{2\ell'+1}} \big( -\frac{\df }{\df \omega} + \frac{\ell'}{2}\cot\frac{\omega}{2} \big) & \ell'= \ell+1
    \end{cases}
    \,.
\end{split}
\end{align}
Terms appearing in the electric Hamiltonian are of the form $\hat{\mathcal{E}}^a_\kappa \hat{\mathcal{E}}^a_{\kappa'}$ (see \cref{eq:HE_max_tree}), with various possible combinations of left and right operators appearing (labels omitted for brevity). 
In terms of spherical components this is
\begin{align}
    \hat{\mathcal{E}}^a_\kappa \hat{\mathcal{E}}^a_{\kappa'} = \hat{\mathcal{E}}^0_\kappa \hat{\mathcal{E}}^0_{\kappa'} - \hat{\mathcal{E}}^+_\kappa \hat{\mathcal{E}}^-_{\kappa'}
    - \hat{\mathcal{E}}^-_\kappa \hat{\mathcal{E}}^+_{\kappa'}
    \,.
\end{align}

Next we compute the matrix elements of the magnetic operators. 
In coordinates, the magnetic operators are just functions of the axis-angle coordinates $(\omega,\theta,\phi)_\kappa$. 
All needed matrix elements can be formed by taking linear combinations and products of the Cayley-Klein parameters i.e. the matrix elements of the four components of $\hat{X}$
\begin{align}
    \bra{\omega'\theta'\phi'}\hat{X}_{{m_L}{m_R}}\ket{\omega\theta\phi} = \delta_{\g \g'}D^{1/2}_{{m_L}{m_R}}(\omega,\theta,\phi)
    \,.
\end{align}
When converting these expressions to the mixed basis, it is useful to decompose $\hat{X}$ into its irreducible representations under transformations generated by $\hat{L}^a$ -- that is, under the transformations
\begin{align}
    \hat{X} \mapsto e^{-i\phi^a \hat{L}^a} \hat{X} e^{i\phi^a \hat{L}^a}
    \,.
\end{align}
Note that $e^{-i\phi^a(\g) \hat{L}^a} = \hat{\Theta}_{L\g}\hat{\Theta}^\dagger_{R\g}$. 
Then using \cref{eq:link_trafo} the transformation can also be written as
\begin{align}
    \hat{X} \mapsto D(\g) \, \hat{X} \, D(\g)^\dagger
    \,,
\end{align}
where the representation matrices $D$ are taken in the defining representation. 
Clearly the trace of $\hat{X}$ is invariant under this transformation, so we identify
\begin{align}
    \hat{S} \equiv \tfrac{1}{2}\Tr \hat{X}
\end{align}
as the scalar component of $\hat{X}$.

The three remaining components of $\hat{X}$ form a vector operator under this transformation. 
To put them in a standard form, we define
\begin{align}
    \hat{X}^a = 2 \Tr T^a \hat{X}
    \,.
\end{align}
These are the Cartesian components of the vector operator. 
The original loop operator is given in terms of its scalar-vector decomposition by
\begin{align}
    \hat{X} = \hat{S} I + \hat{X}^a T^a 
    \,.
\end{align}
For the vector part, it is convenient to instead work with the (covariant) spherical components
\begin{align}
    \hat{X}^0 &= \hat{X}^z \\
    \hat{X}^\pm &= \mp \left( \hat{X}^x \pm i \hat{X}^y \right)
    \,.
\end{align}

The matrix elements of the loop operator can now be expressed simply in the mixed basis. 
The scalar component is independent of all angular quantum numbers and is given by
\begin{align}\label{eq:scalar_mat_elements}
    \bra{\omega' \ell' m'}\hat{S}\ket{\omega \ell m} = \delta_{\ell'\ell}\delta_{m'm}\delta(\omega'-\omega) \cos\frac{\omega}{2}
    \,.
\end{align}
For the vector part, the Wigner-Eckart theorem applies, and we can again write the vector operators $X^q$ using \cref{eq:vec_ops}. 
The reduced matrix elements for the $X$ operator are now
\begin{align}\label{eq:mag_mat_elements}
    \langle \omega'\ell' \vert \vert \bm{X} \vert\vert \omega \ell \rangle = 
    \delta(\omega-\omega')\begin{cases}
    i \sqrt{\frac{\ell}{2\ell'+1}} \sin\frac{\omega}{2} & \ell' = \ell-1 \\
    0 & \ell' = \ell \\
    -i \sqrt{\frac{\ell'}{2\ell'+1}} \sin\frac{\omega}{2} & \ell'=\ell+1
    \end{cases}
    \,.
\end{align}

In two dimensions, terms appearing in the magnetic Hamiltonian are of the form $\Tr\hat{X}_1$, or $\Tr\hat{X}_2^\dagger \hat{X}_1$. 
In terms of the scalar-vector parts, these are
\begin{align}
    \Tr(\hat{X}_1) &= 2\hat{S}_1 \\
    \Tr(\hat{X}_2^\dagger \hat{X}_1) &= 2\hat S_2 \hat S_1 + \tfrac{1}{2}\hat{X}^{a\dagger}_2\hat{X}^a_1.
\end{align}
Similar expressions for three and four loop operators are easily obtained.
Note that for lattices in dimensions three or higher, terms with three and four loop operators also appear. 

Combining \cref{eq:elec_mat_elements,eq:scalar_mat_elements,eq:mag_mat_elements} with the Hamiltonian of the preceding section, one obtains an exact representation of the system. By employing the truncation scheme of \cref{sec:singlePlaquette} on each loop operator, one obtains a truncated representation of the entire system.

\subsection{Dealing with the Global Constraint}
\label{app:GeneralReduction}

The construction so far has removed all gauge redundancies, with the exception of the one arising from a transformation at the origin given by \cref{eq:gauge_trafo_at_origin}.
This remaining transformation is generated by
\begin{align}
    \hat G^a(n_0) = \sum_i \hat{\mathcal{E}}(\ell(n_0, e_i)) = 0
    \,.
\end{align}
Since every loop originates and ends at the origin, one finds (see \cref{app:maxTree}):
\begin{align}
   \hat G^a(n_0) &= \sum_\kappa \left(\mathcal{E}_L^a(\kappa) - \mathcal{E}_R^a(\kappa)\right)
    \nonumber\\
    & = \sum_\kappa -\hat L^a(\kappa)\nonumber\\
    &= -\hat L^a_{\rm tot}
    \,.
\end{align}
This implies that the remaining gauge transformation is enacted by the operator $\hat L_{\rm tot}$ that measures the total charge of the lattice. 
In the absence of charged matter, only the zero charge sector is physical.

In this section, we motivate a general form of the Hamiltonian that is valid in the sector of vanishing total charge, as well as the method for deriving its exact form.
This will be valid
for an arbitrary number of lattice points.
This Hamiltonian will commute with the total charge operator as required. 
\begin{figure}[h]
\centering
\includegraphics[width=0.4\textwidth]{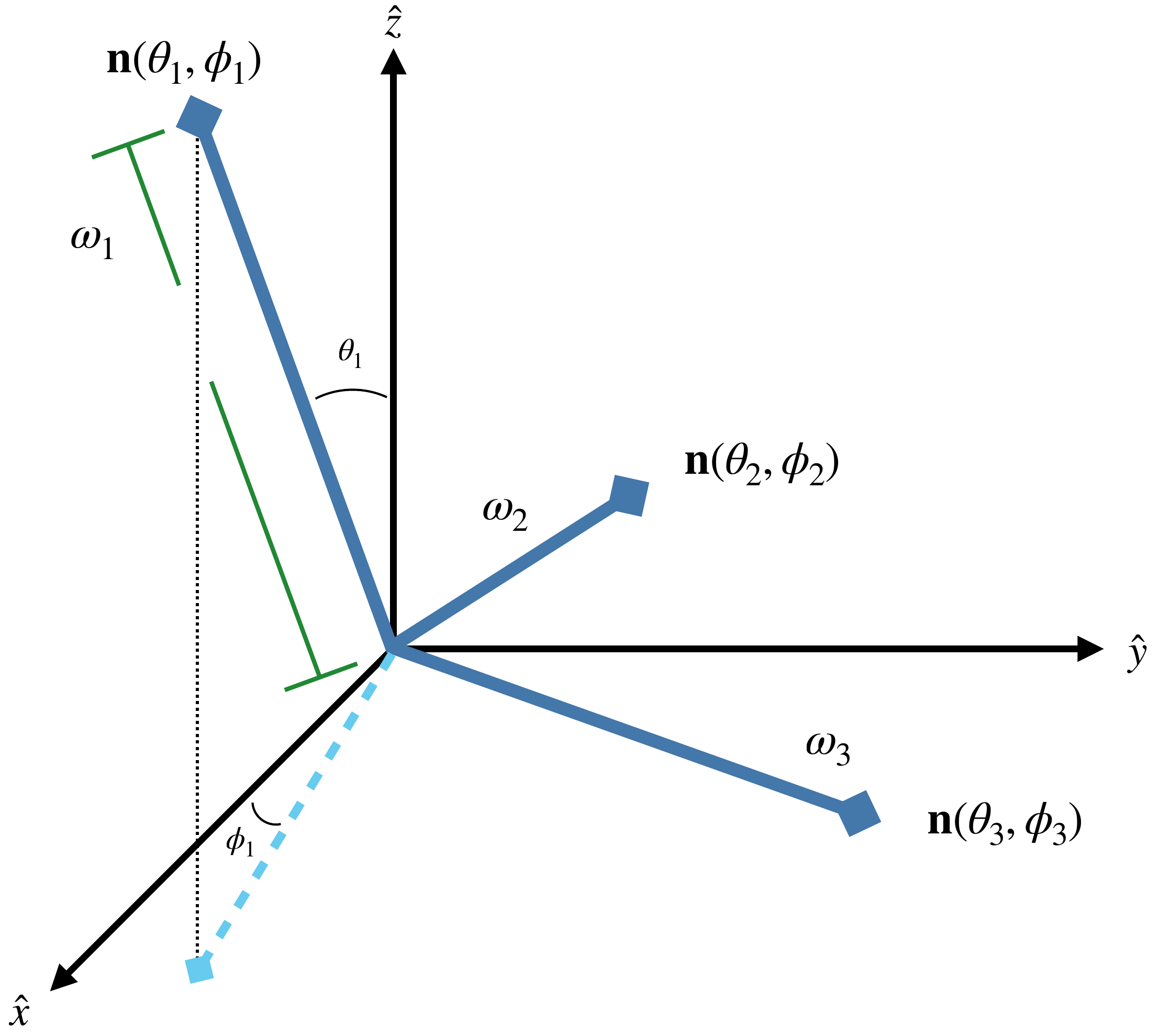}
\caption{Image of rigid rods in 3D space, representing an arbitrary configuration. The angles $\theta_i, \phi_i$ define the orientation of each rigid rod, with length $\omega_i$. }
\label{fig:RigidRods}
\end{figure}

In axis-angle coordinates, each group element is parameterized by three variables: $\omega, \theta$ and $\phi$. 
A classical interpretation of such a configuration is a set of a rigid rods starting at the origin, with length $\omega$ and  polar angles $\theta$ and $\phi$, as shown in \cref{fig:RigidRods}.
The set of $N_L$ rigid rods, each parameterized by $\{\omega_i, \theta_i, \phi_i\}$, define a given `shape'. Here $N_L = 3(n_\ell - N^d + 1)$ is the total number of loop operators that remain after gauge fixing, and where as defined before $n_\ell$ is the total number of links and $N^d$ is the total number of sites. 
The total charge operator $\hat{L}_{\text{tot}}^a$ enacts transformations
\begin{align}
    e^{i\phi^a(\h) \hat{L}^a_{\text{tot}}}\ket{\g_1,...,\g_{N_L}} = \ket{\h\g_1\h^{-1},...,\h\g_{N_L}\h^{-1}}
\end{align}
which act on each loop variable by conjugation. Let $\g = \g(\hat{n},\omega)$, where $(\hat{n},\omega)$ are axis-angle coordinates for $\g$. Note that $(\hat{n},\omega)$ also gives the direction and length of the `rod' representing $\g$. Using the adjoint formula
\begin{align}
    \h \g(\hat{n},\omega) \h^{-1} = \g( R(\h)\hat{n},\omega)
\end{align}
we see that this transformation preserves $\omega$, while rotating $\hat{n}$. Here $R(\h) = D^{1}(\h)$ is the rotation $\h$ in the 1 (vector) representation. 
Then the transformations generated by $\hat{L}^a_{\text{tot}}$ rotate configurations of the rods into other configurations with the same `shape', but a different orientation. 

In the classical interpretation, the total charge represents the total angular momentum of the system, which generates total rotations about the origin and keeps the original shape unchanged.
Such an overall rotation is parameterized by three angular variables, and one therefore expects that this overall constraint removes three angular variables, leaving $3N_L-3$ independent variables. The variables defining a given shape are referred to as ``shape coordinates''.
Since the lengths of the rods, given by the $\omega_i$ are invariant under global rotations, they are always part of the set of all shape coordinates; we will ignore them for the time being. 

In order to motivate the form of the Hamiltonian which separates the energy stored in the shape of the system from that of the overall charge, we must define the shape and orientation coordinates. 
Configurations with the same shape coordinates but different orientation coordinates correspond to different configurations points $P$ on the fiber representing the gauge transformation at the origin, and choosing a particular orientation chooses a representative point $P_0$.
We will refer to the original coordinates as the `axis-angle' coordinates and the new coordinate system as the `physical' coordinates.
Physical coordinates are those parameterizing the representative point $P_0$.

In the `axis-angle' coordinate system, the axes of rotation are given by
\begin{align}
\tilde{\bm n}_i = \{\sin \theta_i \cos \phi_i, \sin \theta_i \sin \phi_i, \cos \theta_i\}
\,.
\end{align}
As discussed, the `physical' coordinate system is obtained by fixing the orientation of the shape of the $N_L$ rods. 
The `axis-angle' coordinate system is then recovered from the physical coordinate system by carrying out a global rotation on the whole rigid system. 

To pick a representative point $P_0$ corresponding to a gauge fixing convention, we choose to lock the first two rotors with an angle of $\theta_{12}$ between them and have both of them lying in the $\hat x-\hat z$ plane, with the $\hat z$ axis bisecting the opening angle. 
All the other rotors will then have their axis defined in relation to this plane. 
Therefore, the initial orientation of each rotor is given by
\begin{align}
\bm n_1^{(0)} &= \left(-\sin \frac{\theta_{12}}{2},0,\cos \frac{\theta_{12}}{2}\right) \nonumber \\
\bm n_2^{(0)}&=\left(\sin \frac{\theta_{12}}{2},0,\cos \frac{\theta_{12}}{2}\right) \nonumber \\
\bm n_\mu^{(0)} &= \{\sin \theta_{0\mu} \cos \phi_{0\mu}, \sin \theta_{0\mu} \sin \phi_{0\mu}, \cos \theta_{0\mu}\}
\,,
\end{align}
where $\mu$ ranges from 3 to $N_L$ and we use the subscript $\{0\mu\}$ to denote that $\theta_{0\mu}$ and $\phi_{0\mu}$ are the angles between the $\mu^\text{th}$ rods and the vector that is the sum of the axes of the first two rods. 
Note that this is not a unique definition of $\bm n_i$, but we find it particularly convenient. 
These $2 (N_L-2)+1$ angles (plus the $\omega$ variables, of which there are $N_L$) are the shape variables on which the wave function will depend; we will collectively refer to the set of shape variables as $\bm S$. 

We now perform a simultaneous rotation on this entire system using three Euler angles, $\alpha, \beta, \gamma$; these are the three orientation variables, which we will refer to as $\bm O_i$. 
In particular, in this coordinate system, the $N_L$ axes are given by
\begin{align}
\bm n_i &= R_z(\alpha)R_y(\beta)R_z(\gamma)\bm n^{(0)}_i \nonumber \\
&= \CR (\alpha, \beta, \gamma) \bm n^{(0)}_i
\,,
\label{eq:RotDef}
\end{align}
where $R_{z}(\theta)$ and $R_{y}(\theta)$ is defined as the rotation around the $\hat z$ and $\hat y$ axes, respectively, by angle $\theta$. 
For example, 
\begin{align}
R_z(\theta) = \left(
\begin{array}{ccc}
 \cos \theta  & -\sin \theta  & 0 \\
 \sin \theta  & \cos \theta  & 0 \\
 0 & 0 & 1 \\
\end{array}
\right)
\,.
\end{align}

In order to rewrite the Hamiltonian in the `physical' coordinate bases, the variables in the `axis-angle' coordinates need to be related to those in the `physical' coordinate system. 
This can be done by simply equating $\bm n_i =\tilde{\bm  n}_i$. 
Doing so, we find that the shape variables are related to the `axis-angle' variables via
\begin{align}
\cos \theta_{12} = &\cos \theta_1 \cos \theta_2 + \cos\left(\phi_1 - \phi_2\right) \sin \theta_1 \sin \theta_2 \nonumber \\
\cos \theta_{0\mu}=&\frac{1}{2\cos \left(\frac{\theta_{12}}{2}\right)} \big[ \cos \theta_{\mu} (\cos \theta_1+\cos\theta_2)+\sin \theta_{\mu}\nonumber\\
& \times \left(\sin \theta_1 \cos (\phi_1-\phi_{\mu})+\sin \theta_2 \cos (\phi_2-\phi_{\mu})\right)\big]\nonumber \\
\cos \phi_{0\mu}=&\frac{1}{2 \sin \theta_{0\mu}\sin \frac{\theta_{12}}{2}}\big[\cos \theta_{\mu}\left(\cos \theta_1-\cos \theta_2\right)+ \sin \theta_{\mu}\nonumber\\
& \times \left(\sin \theta_1 \cos (\phi_1 - \phi_{\mu})-\sin \theta_2 \cos (\phi_2 - \phi_{\mu})\right)\big]
\,,
\end{align}
while the three orientation variables are related to the original variables via
\begin{align}
\cos \beta &=\frac{\cos \theta_1 + \cos \theta_2}{2 \cos \frac{\theta_{12}}{2}}\nonumber \\
\tan \alpha &= \frac{\sin \theta_1 \sin \phi_1+ \sin \theta_2 \sin \phi_2}{\sin \theta_1 \cos \phi_1 + \sin \theta_2 \cos \phi_2} \nonumber \\
\tan \gamma &= -\frac{\sin \theta_1 \sin \theta_2 \sin (\phi_1 - \phi_2)}{\left(\cos \theta_1- \cos \theta_2\right)\cos \frac{\theta_{12}}{2}}
\,.
\end{align}

The goal is now to carry out the change of coordinate system from the axis-angle to the physical basis. 
This requires making use of the relations
\begin{align}
\pdv{F[\bm X]}{X_i} =&\left(\pdv{\tilde X_j}{X_i}\right)\pdv{F[\tilde{\bm X}]}{\tilde X_j} \nonumber \\
\pdv{F[\bm X]}{X_i}{X_j} =&\left(\pdv{\tilde X_k}{X_i}{X_j}\right) \pdv{F[\tilde{\bm X}]}{\tilde X_k}\nonumber \\
&+\left(\pdv{\tilde X_k}{X_i}\right)\left(\pdv{\tilde X_l}{X_j}\right)\pdv{F[\tilde{\bm X}]}{\tilde X_k}{\tilde X_l}
\,,
\end{align}
where $\bm X = \{\theta_i, \phi_i\}$ and $\tilde{\bm X} = \{\theta_{12}, \theta_{0\mu}, \phi_{0\mu}, \alpha, \beta, \gamma\}$.

Since the generators of Gauss' law are equal to the total angular momentum of the system, they only depend on the orientation coordinates parameterizing the rotation of the system. 
And indeed, when the change of coordinates is carried out for these operators, we find that 
\begin{align}
\hat L_\text{tot}^x &= i \left(\sin \alpha \pdv{}{\beta}+ \cot \beta \cos \alpha \pdv{}{\alpha}- \csc \beta \cos \alpha \pdv{}{\gamma}\right) \nonumber \\
\hat L_\text{tot}^y &= i \left(-\cos \alpha \pdv{}{\beta}+ \cot \beta \sin \alpha \pdv{}{\alpha}- \csc \beta \sin \alpha \pdv{}{\gamma}\right)\nonumber \\
\hat L_\text{tot}^z &= -i  \pdv{}{\alpha} 
\,, 
\end{align}
regardless of the size of the lattice.

Since the Hamiltonian of the system commutes with the Gauss' law generators, the most general form for a (lattice gauge theory) Hamiltonian is given by
\begin{align}
\hat H =& \hat H_B(\bm S) + \hat H_E(\bm S, \bm O;\partial_{\bm S},\partial_{\bm O}) \\
=&\hat H_B(\bm S) +\hat H^{[0]}_E(\bm S; \partial_{\bm S})+ \hat H_E^{[\bm L_\text{tot}]} (\bm S, \bm O;\partial_{\bm S},\partial_{\bm O}) \,  \nonumber
\label{eq:HamGeneralForm}
\end{align}
where $\hat H_E^{[\bm L_\text{tot}]}$ is given by
\begin{align}
\hat H_E^{[\bm L_\text{tot}]} &\equiv \eta(\bm S) \hat L^2_\text{tot} + v^a (\bm S, \bm O) \hat L_\text{tot}^a \\
&+  t^{ab} (\bm S, \bm O)\big\{\hat L_\text{tot}^a, \hat L_\text{tot}^b\big\}+\tilde{v}^a_{\bm S_i}(\bm S, \bm O) \pdv{}{\bm S_i}\hat L_\text{tot}^a \nonumber \, .
\end{align}

In this expression, $\eta, v^a, \tilde v^a$ and $t^{ab}$ are a scalar, two vectors and a traceless symmetric (rank two) tensor, respectively and the index $i$ is summed over all shape variable in the last term of the second expression.
This form is motivated by the observation that $\hat \CE_{L/R}^a$ are all first order differential operators and therefore $\hat H$ can be at most bilinear in $\hat L_\text{tot}^a$. 

It is trivial to see that both $\hat H_B$ and $\hat H^{[0]}_E$ commute with any of the angular momentum operators $\hat L_\text{tot}^a$, as the operators $\hat L_\text{tot}^a$ only depend on orientation variables and their conjugate momenta while $\hat H_B$ and $\hat H^{[0]}_E$ only depend on shape variables and their conjugate momenta. 
It is easy to guarantee that the first term of $\hat H_E^{[\bm L_\text{tot}]}$ commutes with $\hat L_\text{tot}^a$, as it simply requires that the coefficient $\eta(\bm S)$ only depend on the shape variables,
\begin{align}
[\hat L_\text{tot}^a, \eta(\bm S)] = 0 \, .
\end{align}
The rest of  $\hat H_E^{ L_\text{tot}}$ commutes with $\hat L_\text{tot}^a$ as long as the vectors $v^a$ and $\tilde{v}^a$, as well as symmetric traceless tensor $t^{ab}$ satisfy the commutation relations
\begin{align}
\left[\hat L_\text{tot}^a, V^b\right] = i f^{abc}V^c\, , \qquad V = v, \tilde v
\,,
\label{eq:VecCom}
\end{align}
and
\begin{align}
\left[\hat L_\text{tot}^a,t^{bc}\right] & = i \left(f^{abd}\delta^{c e} - f^{ad c}\delta^{be}\right)t^{d e} \nonumber\\
&=i \left(f^{abd}t^{dc} + f^{acd}t^{db}\right)
\,.
\label{eq:TensorCom}
\end{align}
These commutation relations greatly restrict the form of the vectors and tensors. 
In particular, a vector $V$ that satisfies \cref{eq:VecCom} must be of the form
\begin{align}
V^a &= \CR^{ab} (\alpha, \beta, \gamma) \nu^b (\bm S) 
\,,
\label{eq:VecGen}
\end{align}
where $\CR (\alpha, \beta, \gamma)$ is defined in \cref{eq:RotDef} and $\nu^b(\bm S)$ is an arbitrary function of the shape variables. Furthermore any tensor that satisfies \cref{eq:TensorCom} can be written as
\begin{align}
t^{ab}=&\Bigg(\frac{\CR^{ac}(\alpha, \beta, \gamma)\CR^{bd}(\alpha, \beta, \gamma)}{2}\nonumber\\
&+\frac{\CR^{ad}(\alpha, \beta, \gamma)\CR^{bc}(\alpha, \beta, \gamma)}{2}
\nonumber\\
&- \delta^{ab}\frac{\CR^{ec}(\alpha, \beta, \gamma)\CR^{ed}(\alpha, \beta, \gamma)}{3}\Bigg)\mu^c (\bm S) \nu^d (\bm S)\, .\nonumber
\end{align}
Since an arbitrary traceless symmetric rank two tensor can always be constructed out of two vectors as
\begin{align}
\label{eq:TensorCriteria}
T^{ab} =& \frac{U^a V^b + U^b V^a}{2}- \frac{1}{3}\delta^{ab}U^c V^c \, ,
\end{align}
this implies that $t^{ab}$ is constructed out of vectors that satisfy \cref{eq:VecGen} ie
\begin{align}
U^a &= R^{ab}(\alpha, \beta, \gamma)\mu^b(\bm S)\nonumber \\
V^a &= R^{ab}(\alpha, \beta, \gamma)\nu^b(\bm S)
\end{align}
These results can be easily shown given the commutation relation of the rotation operator $\CR(\alpha, \beta, \gamma)$ with the angular momentum operator:
\begin{align}
\left[\hat L_\text{tot}^a,\CR^{bc}\right] & = i f^{abd}\CR^{d c}
\,.
\end{align}
Once the Hamiltonian is written in this form, it is trivial to work in the physical (charge zero) Hilbert space. 
The Hamiltonian that is to be simulated is simply
\begin{align}
\hat H &= \hat H_B(\bm S) + \hat H^{[0]}_E(\bm S; \partial_{\bm S})
\,,
\end{align}
which can be obtained by using the variable transformation given above and dropping all dependence on the orientation variables and their derivatives.

The practical utility of this coordinate transformation, particularly for larger lattices, depends on how computationally expensive it is to carry out the algebraic manipulations necessary to calculate $H_E^{[0]}$. 
We have explicitly done this for a two-plaquette system, but this question will be explored in greater detail in follow-up work.

\section{Conclusions}\label{sec:conclusions}

The ability to carry out \emph{ab-initio} calculations of the properties of gauge theories, particularly time-dependent behavior, without resorting to perturbation theory would provide an invaluable took for comparing such theories to experimental results.  
A promising approach to this problem, which has recently seen a surge of interest, is the simulation of Hamiltonian formulations of  lattice gauge theories on digital quantum computers.
In these formulations, the  finite lattice spacing and finite volume regulate the UV and IR divergences, respectively, of the underlying continuum theory. 
Unfortunately, for continuous gauge groups, such as those in the Standard Model  of particle physics, the Hilbert spaces spanned by the pertinent Hamiltonians are  still infinite dimensional and therefore must be truncated before the theory can be numerically simulated.

A key challenge is finding a finite-dimensional Hilbert space that faithfully reproduces key properties of the theory while requiring a practical number of quantum resources.
In particular, while the theory is relatively simple to truncate at strong bare coupling, these truncations, which have been used widely in the literature, become inefficient at weak bare coupling. 
This inefficiency is especially concerning as weak bare coupling is required to ultimately take the continuum limit of the lattice gauge theory; however it remains to be seen how small of a bare coupling is really required to obtain the continuum limit reliably.
The main finding of this work is a novel method for carrying out this truncation that works well both at strong and weak bare coupling. 
Our approach utilizes the maximal tree method for gauge-fixing the theory, which automatically enforces Gauss' law at all sites.
This gauge fixing results in a Hamiltonian whose Hilbert space is a tensor product of mostly unconstrained (up to the global gauge transformation) \sutwo group elements.
Each of these group elements is then represented in a mixed axis-angle  basis, with a continuous label $\omega$ representing the rotation angle, and spherical harmonics characterized by integers $\ell, m$ representing the angular distribution of the rotation axis.

In this basis, the lowest lying wave functions are always slowly varying, such that they are dominated by small $\ell$ values. 
The dependence on $\omega$, however, depends on the value of the bare gauge coupling: at large coupling the dependence on $\omega$ is also slowly varying while for weak coupling the low-lying wave functions become localized around $\omega_i = 0$. This localization at weak coupling means that the low-lying wavefunctions are exponentially suppressed for $\omega > \omega_{\max}$, where $\omega_{\max}$ scales linearly with the bare gauge coupling $g$. 
A truncated Hilbert space can therefore be obtained by truncating the allowed $\ell$ values, and sampling the dependence on $\omega$ at discrete points $\omega_n$  in the range $0 < \omega < \omega_{\max}$. 
While we have not provided detailed numerical studies, we have shown that this indeed works well numerically for a system with a single plaquette.
A detailed numerical study, as well as a construction of an explicit quantum circuit is left for future work; of particular interest is how the circuit depth scales with relevant parameters.
Additionally, it is important to note that maximal-tree gauge fixing is an  incomplete gauge-fixing procedure and there remains a global gauge transformation. This implies that the Hamiltonian still spans all color charge sectors.  However, we have demonstrated how this theory can be fully gauge-fixed in theory, resulting in a Hamiltonian that spans only the physical (colorless) sector; the practical application of this method is left for future work. 

While some of our discussion has assumed a two-dimensional lattice, the results presented are general and work in arbitrary dimensions.
Additionally, this work has focused on an \sutwo gauge theory; however we believe that this can be generalized to more complicated non-Abelian groups, in particular the important case of \suthree. This is again left for future work. 

\begin{acknowledgments}
We would like to thank Ivan Burbano and Christopher Kane for enlightening discussions on the topic.
We are also grateful to Ivan Burbano, Hank Lamm for comments on the manuscript.
This work was supported by the U.S. Department of Energy (DOE), Office of Science under contract DE-AC02-05CH11231, through Quantum Information Science Enabled Discovery(QuantISED) for High Energy Physics (KA2401032), by the Office of Advanced Scientific Computing Research (ASCR) through the Accelerated Research for Quantum Computing Program, by the U.S. Department of Energy grant DE-FG02-97ER-41014 (Farrell), the U.S. Department of Energy, Office of Science, Office of Nuclear Physics and \href{https://iqus.uw.edu}{\color{black}}{InQubator for Quantum Simulation (IQuS)} under Award Number DOE (NP) Award DE-SC0020970, and the U.S. National Science Foundation under grant PHY1316222. This work is also supported by the U.S. Department of Energy, Office of Science under contract DE-AC02- 05CH11231 and through the \href{https://phys.washington.edu}{\color{black}}{Department of Physics} and \href{https://www.artsci.washington. edu}{\color{black}}{the College of Arts and Sciences}  at the University of Washington. This work was performed in part at the Munich Institute for Astro-, Particle and BioPhysics (MIAPbP) which is funded by the Deutsche Forschungsgemeinschaft (DFG, German Research Foundation) under Germany´s Excellence Strategy – EXC-2094 – 390783311 and at the Aspen Center for Physics, which is supported by National Science Foundation grant PHY-2210452.
\end{acknowledgments}

\appendix

\section{Electric Bases}\label{app:elec_bases}
This section concerns bases for the one-link Hilbert space that diagonalize the electric operators. As mentioned in \cref{sec:basics_KG}, an electric basis is one that diagonalizes the $\hat E^2 = \hat E_L^2 = \hat E_R^2$ operator.
This leaves freedom to specify which additional mutually-commuting operators that commute with $\hat E^2$ are chosen to be diagonalized.

A common choice is the set$\hat{E}^2, \hat{E}_L^z$, and $\hat{E}^z_R$, with eigenvalues
\begin{align}
\label{eq:Eop_electric_basis_app}
    \hat E^2 | j {m_L}{m_R} \rangle &= j(j+1)| j {m_L}{m_R} \rangle
    \nonumber\\
    \hat E^z_L | j {m_L}{m_R} \rangle &= m_L | j {m_L}{m_R} \rangle
    \nonumber\\
    \hat E^z_R | j {m_L}{m_R} \rangle &= m_R | j {m_L}{m_R} \rangle
    \,.
\end{align}

This is the original basis used by Kogut and Susskind, picked due to simplicity and its utility for performing a strong coupling perturbative expansion.

Another choice of electric basis that arises naturally when one works in axis-angle coordinates is the $\ket{j \ell m}$ basis, which uses the operator
\begin{align}
    \hat L^a \equiv \hat E^a_R - \hat E^a_L\,.
\end{align}
In this basis, the operators $\hat{E}^2$, $\hat{L}^2$, and $\hat{L}^z$ are diagonal, with
\begin{align}
        \hat E^2 | j \ell m \rangle &= j(j+1)| j \ell m \rangle
    \nonumber\\
    \hat L^2 | j \ell m \rangle &= \ell(\ell+1) | j \ell m \rangle
    \nonumber\\
    \hat L^z | j \ell m \rangle &= m | j \ell m \rangle
    \,.
\end{align}
For fixed $j$, the range of quantum numbers is $\ell=0,\hdots,2j$, $m=0,\hdots,\ell$, for total degeneracy $(2j+1)^2$.

The relation between these bases is~\cite{khersonskii1988quantum}
\begin{align}
\label{eq:Uexp}
    \ket{j{m_L}{m_R}} & = \sum_{\ell,M}(-i)^\ell \frac{2 \ell+1}{2j+1} C^{j{m_R}}_{j{m_L}\ell M}
    \ket{j \ell M}
    \,,
\end{align}
where $C^{jm'}_{jm\ell M}$ is a Clebsch--Gordan coefficient. In axis-angle coordinates, up to normalization the states are
\begin{align}
    \ket{j \ell m} = \chi^j_\ell(\omega)Y^{\ell}_m(\theta, \phi)
    \,,
\end{align}
where $\chi^j_\ell(\omega)$ is the generalized character of order $\ell$ of the irrep of rank $j$, which satisfy the differential equation
\begin{align}
\label{eq:diffEq_chi}
    \left[\frac{{\rm d}^2}{{\rm d} \omega^2}+\cot\frac{\omega}{2} \frac{\rm d}{{\rm d} \omega}+j(j+1) - \frac{\ell(\ell+1)}{4\sin^2\frac{\omega}{2}}\right]\chi^j_\ell(\omega) = 0
    \,.
\end{align}
Note that this is just the one-plaquette radial equation, \cref{eq:single_ham}, at infinite coupling. %
The generalized characters are orthogonal and complete
\begin{align}
    \int_0^{2\pi}\! \dd \omega \, 4 \sin^2\frac{\omega}{2} \,\,\chi^{j_1}_\ell(\omega)\, \chi^{j_2}_\ell(\omega) &= 4 \pi \,\delta_{j_1 j_2}\nonumber\\
    \sum_{j,\ell} \chi^{j}_\ell(\omega_1)\, \chi^{j}_\ell(\omega_2) &= \frac{\pi \delta(\omega_1 - \omega_2)}{\sin^2\frac{\omega}{2}}\,.
\end{align}

\section{One-Plaquette Universe: Analytic Solution at All Coupling}
\label{app:analytic_solution}
The energies and energy eigenstates of the one-plaquette system studied in \cref{sec:singlePlaquette} can be expressed exactly in terms of certain special functions. 
These exact solutions are useful for analyzing the numerical results, as in \cref{fig:one_plaquette_HB}. In this section we show how to obtain these exact solutions.

We begin with the Hamiltonian of \cref{eq:one_plaq_ham}
\begin{align}
        \hat H = \frac{2g^2}{a}\hat{\cal E}^2 + \frac{2}{g^2 a}\left(1 - \cos\frac{\omega}{2}\right)
    \,,
\end{align}
which possesses no local gauge transformations, but still has one global gauge transformation.

Viewed as a differential equation in axis-angle coordinates, the time-independent Schrodinger equation $H\ket{\Psi} = \lambda \ket{\Psi}$ is separable into `radial' coordinate $\omega$ and `angular' coordinates $(\theta,\phi)$. 
Using the ansatz $\Psi = R(\omega)Y(\theta,\phi)$, the equation separates into
\begin{align}
    L^2Y = \ell(\ell+1)Y
\end{align}
and
\begin{align}
    \left[ -\frac{\dd ^2}{\dd \omega^2} - \cot\frac{\omega}{2}\frac{\dd }{\dd \omega} + \frac{\ell(\ell+1)}{4\sin^2\frac{\omega}{2}} + \frac{1}{2g^2}V(\omega)  \right]R = \frac{\lambda }{2g^2}R
    \,,
\end{align}
where the operator $L^2$ is the spherical Laplacian which has as its eigenfunctions the spherical harmonics $Y_{\ell m}(\theta,\phi)$.
The volume form being proportional to $\sin^2\frac{\omega}{2}$ motivates a change of variables $u(\omega)=R(\omega) 2 \sin\frac{\omega}{2}$, such that the differential equation becomes
\begin{align}
    u'' + \left[\frac{1}{4} - \frac{1}{g^4}\left(1-\cos\frac{\omega}{2}\right) + \frac{\lambda}{2g^2} + \frac{\ell(\ell+1)}{4 \sin^2 \frac{\omega}{2}} \right]u = 0
    \,.
\end{align}
For $\ell=0$ this equation has a solution in terms of Mathieu functions. 
Incidentally, only the $\ell=0$ states are physical (gauge invariant) states, so we content ourselves with this case. 
In terms of the re-scaled radial function $u$, the $\ell=0$ radial equation is
\begin{align}
    u'' + \left[\frac{1}{4} - \frac{1}{g^4}\left(1-\cos\frac{\omega}{2}\right) + \frac{\lambda}{2g^2}\right]u = 0
    \,,
\end{align}
which is a Mathieu equation. 
The energies $\lambda$ are determined by the boundary conditions $u(0)=0$, $u(2\pi)=0$.

Further setting $z=\omega/4$, we get
\begin{align}
    u''(z) + \left[4 + \frac{8\lambda}{g^2} - \frac{16}{g^4} + \frac{16}{g^4}\cos(2z)\right]u(z)=0
    \,.
\end{align}
Comparing to the standard form of the Mathieu differential equation $y''+(a-2q\cos(2z))y=0$, we have the relations
\begin{align}
    a&=4 + \frac{8\lambda}{g^2} - \frac{16}{g^4} \nonumber\\
    q&=-8/g^4
    \,.
\end{align}
The Mathieu differential equation
\begin{align}
    y''+(a-2q\cos(2z))y=0
\end{align}
is a linear second order differential equation with even solutions denoted $C(a,q,z)$ and odd solutions $S(a,q,z)$. 
For nonzero $q$, the functions only become periodic for certain special values of $a$, called the \textit{characteristic values}. 
For a fixed non-zero $q$, there is a sequence of such characteristic values $a_n(q)$, and $b_n(q)$ for $n=1,2,3\hdots$. 
The corresponding periodic functions are denoted $\text{ce}_n(q,z)$ and $\text{se}_n(q,z)$, respectively. They are also sometimes called cosine-elliptic and sine-elliptic functions, or Mathieu functions of the first kind.

The functions $\text{ce}_n(q,z)$ and $\text{se}_n(q,z)$ can be categorized by their parity and periodicity as follows
\begin{center}
\begin{tabular}{c|c|c}
    Function \quad & \quad Parity \quad  & \quad Period \\
    \hline
    $\text{ce}_n$, $n$ even & Even & $\pi$ \\
    $\text{ce}_n$, $n$ odd & Even & 2$\pi$ \\
    $\text{se}_n$, $n$ even & Odd & $\pi$ \\
    $\text{se}_n$, $n$ odd & Odd & $2\pi$
\end{tabular}
\end{center}

The radial function $R(\omega)$ for $\ell = 0$ are even in $\omega$. 
This means that the function $u(\omega)$ has to be odd, since it is divided by $\sin(\omega /2)$. 
Since we want the function to be $\pi$ periodic in $z$, this implies that the functions $u(\omega)$ are given by
\begin{align}
    u(\omega) = N \text{se}_n(q,z), \qquad n=2,4,6...
    \,,
\end{align}
where $n$ is an even integer and $N$ is a normalization constant. The energy eigenvalues are given by
\begin{align}
    \lambda_n = \frac{2}{g^2} + \frac{g^2}{8} \left( b_n(q) - 4\right)
    \,,
\end{align}
where again $n$ is an even integer and $b_n$ denotes the Mathieu characteristic numbers. 
In particular, the ground state is 
\begin{align}
    \Psi_0(\omega,\theta,\phi) &= N \frac{\text{se}_2(q,\omega/4)}{\sin(\omega/2)}, \nonumber\\
    q &= -8/g^4 \nonumber
    \,,
\end{align}
where $N$ is a normalization constant.

\section{Review of the Maximal Tree Construction}
\label{app:maxTree}

Before gauge-fixing, the Kogut-Susskind Hamiltonian has as its Hilbert space the tensor product of the space $L^2(G,\dd\g)$ for each link of the lattice. 
Here $G$ is a compact gauge group, and $\dd\g$ is the Haar measure with some normalization. 
For \sutwo the volume of the group is given by $|G| = \int \! \dd\g = 16\pi^2$.
If each of these configuration spaces is parameterized by some coordinates $\phi^a \equiv \phi^a(\g)$, then general states can be thought of as a wave function $\Psi(\{\phi^a_\ell\})$ over those parameters on each link. 
Gauge fixing constructions involve making a change of coordinates $\{\phi^a_\ell\} \mapsto \{y_j,x_k\}$ to a set of coordinates $x_k$ which are gauge invariant, and a set $y_j$ which purely move between gauge-equivalent configurations and/or into the unphysical Hilbert space. 
Then any wave function in the zero-charge sector, being gauge invariant, will depend only on the variables $x_k$. 
In such a sector, the gauge-orbit variables $y_j$ may all be fixed to any desired value, typically those associated with the identity element of the group.

In this viewpoint, it is clear that in order to re-express the Hamiltonian as a differential operator with respect to the variables $x_k$, one need only use the chain rule on the change of coordinates $\{\phi^a_\ell\} \mapsto \{y_j,x_k\}$. 
This is exactly the approach taken in~\cite{PhysRevD.31.2020, MULLER198449}.
On the other hand, the approach taken by ~\cite{Ligterink_2000} leads to a cleaner procedure for expressing the new electric operators in terms of the old electric operators, which we now summarize.

\subsection{Maximal Tree and Loop Variables}

Consider a $d$-dimensional lattice containing $N^d$ sites labelled by $n=(n_1, \ldots n_d)$, and with links labelled by $\ell=(n,i)$. Here  $i=1,\ldots, d$ labels the dimension, and for each dimension, $e_i$ denotes the corresponding unit vectors. 
Each link gets a copy of the $L^2(G,\dd\g)$ Hilbert space, and can be labeled by a group element $\g$. 
Following the discussion in \cref{sec:basics_KG} one can now define a link operator $\hat U(n,e_i)_{mn}$ and electric operators $\hat E^a_{L}(n,e_i)$ and $\hat E_R^a(n,e_i)$ generating left and right gauge transformations on link $(n,i)$ respectively. 
For each lattice site, the commutation relations between these operators are those given in \cref{eq:EEcommutators,eq:EUcommutators}, and operators at different lattice sites commute with one another.
In the remainder of this section we will often restrict ourselves to two-dimensional lattices, such as that shown in \cref{fig:2dLattice}, but all results can easily be generalized to a general $d$-dimensional lattice. 

Next one defines a certain collections of links on this lattice. A tree is a collection of links with no closed loops, and a maximal tree is a tree that is no longer a tree if any additional links are added to it. 
There are many different choices possible for a maximal tree, and this work chooses the maximal tree in two dimension as the set of links in the $x$ directions for $y =0$, as well as all links in the $y$ direction. 
The maximal tree used is illustrated in \cref{fig:2dLattice} by the collection of red dotted links. In a three-dimensional lattice one would choose the links in the $x$ direction for $y = z = 0$, the links in the $y$ direction for $z = 0$ and all links in the $z$ direction. 
This is the maximal tree definition used by \cite{MULLER198449, PhysRevD.31.2020, Ligterink_2000}. 

There is always enough gauge freedom to `eliminate' the degrees of freedom on the link in a maximal tree~\cite{PhysRevD.15.1128}, leaving the degrees of freedom on the links that are not on the maximal tree. 
We will call such links physical links; in \cref{fig:2dLattice}, they are denoted in green solid lines.  
To keep track of the gauge transformations, one defines loops that originate and end at a fixed point on the lattice (called the origin and chosen as the lattice point with $n_0=(0,0)$ in our example), such that one can identify each physical link $\kappa$ with a loop $P(\kappa)$. 
There are many different loop conventions possible, but for concreteness we will use the choice made by \cite{Ligterink_2000}. 
In this convention, for each physical link $\kappa$ one chooses a path $P(\kappa)$ that goes from the origin of the lattice to the starting point of the physical link traversing only links on the maximal tree, then passes through the physical link, and returns back to the origin traversing only links on the maximal tree. 
An example path $P(\kappa)$ is also shown in \cref{fig:loop1}.

A loop operator $\hat X(\kappa)$ is now defined as the oriented product of link operators along the path $P(\kappa)$
\begin{align}
    \hat X(\kappa) = \prod_{\ell \in P(\kappa) } \hat U(\ell)^{\sigma_\ell}
    \,,
\end{align}
where $\sigma_\ell$ is $+1$ if the link $\ell$ is traversed in the positive orientation, and $-1$ if it is traversed in the negative orientation. 

It will also prove useful to define the path $ {\cal P}(n \to n')$ as the path from the site $n$ to the site $n'$ along the maximal tree,  as well as
the operator
\begin{align}
    \hat W(n', n, {\cal P}) = \prod_{\ell \in {\cal P}(n,n')} \hat U(\ell)^{\sigma_\ell}
    \,.
\end{align}
This allows the loop variable to be written as
\begin{align}
    \hat X(\kappa(n,i)) = \hat W(n) \, \hat U(n,i) \,\hat W(n+e_i)^\dagger
    \,,
\end{align}
where $\hat W(n)$ and $\hat W(n+e_i)$ are shorthand for $\hat W(n, n_0, {\cal P})$ and $\hat W(n+e_i, n_0, {\cal P})$, respectively.

Since $\hat X(\kappa)$ is a product of link operators that start end at the origin, it depends only on gauge transformations made at the origin
\begin{align}
\label{eq:global_gauge_y_app}
    \hat X(\kappa) \to \Omega(n_0)\hat X(\kappa) \Omega(n_0)^\dagger. 
\end{align}
Clearly $\Tr[\hat X(\kappa)]$ is gauge invariant.
%

\subsection{Relation of the Plaquette Operators to the Loop Operators}

Each plaquette is comprised of four links, at least one of which has to be a physical link. 
For a two-dimensional lattice the maximum number of physical links for each plaquette is two, while in higher dimension one can have up to four physical links per plaquette. 
In the example given in \cref{fig:2dLattice} the plaquettes in the bottom row contain a single physical link, while all others contain two physical links. 
Given the set of physical links, the plaquette operator can be written in terms of the loop operators as
\begin{align}
\label{eq:P_through_X}
    \Tr \hat P_p = \Tr \prod_{\kappa\in p} \hat X(\kappa)^{\sigma(\kappa)}
    \,,
\end{align}
where the product runs over the physical links of the plaquette, the operators $\hat X(\kappa)$ are the loop operators corresponding to each physical link, and $\sigma(\kappa)$ is $+1$ $(-1)$ if the link $\kappa$ is traversed by the plaquette in the positive (negative) direction. 
Note that only the trace of the two sides are equal, since the $\hat X$ operators are parallel transported to the origin.

\subsection{Action of Electric Operators on Loop Operators}

In this section the action of the electric operators $\hat E_{L}^a({\ell})$ on the loop variables $\hat X(\kappa)$ will be determined, where $\ell$ denotes any link on the lattice.
As illustrated in \cref{fig:loop1}, there are several cases:
%

\subsubsection*{Case 1: The electric operator \texorpdfstring{$E_L^a(\ell)$}{Ea(l)} is located on a link that is not part of the loop \texorpdfstring{$P(\kappa)$}{P(kappa)}. }
In this case 
\begin{align}
    [\hat E_L^a(\ell), \hat X(\kappa)] = 0
\end{align}
because operators on different links commute.

\subsubsection*{Case 2: The electric operator \texorpdfstring{$E_L^a(\ell)$}{Ea(l)} is located on a link that is part of the outgoing sub-path of loop \texorpdfstring{$P(\kappa)$}{P(kappa)}. }
In this case, we compute, being verbose:
\begin{align}
    & [\hat E_L^a (\ell(n',e_{i'})), \hat X(\kappa(n,e_i))] \\
 & \quad = \hat W(n')[\hat E_L^a (\ell(n',e_{i'})),\hat U(n',e_{i'})] \nonumber\\
 & \quad \quad \times \hat W(n'+e_{i'} \to n) \hat U(\kappa(n,\mu))\hat W^\dagger(n+\mu)
 \,,\nonumber
\end{align}
where $\hat W(n'+e_{i'} \to n)$ denotes the product of link operators along the path from $n'+e_{i'}$ to $n$
\begin{align}
    \hat W(n'+e_{i'} \to n) = \hat W(n) \hat W^\dagger(n'+e_{i'})
    \,.
\end{align}
Using the commutator between $\hat E_L^a(\ell)$ and $\hat U(\ell)$ one finds
\begin{align}\label{eq:case2}
    & [\hat E_L^a (\ell(n',i')), \hat X(\kappa(n,i))] \\
    & \quad =  \hat W(n')T^a \hat W(n' \to n) \hat U (\kappa(n,i))\hat W^\dagger(n+i) \nonumber\\
    & \quad = \left[ \hat W(n')T^a \hat W^\dagger(n') \right] \hat X(\kappa(n,i))
    \,,\nonumber
\end{align}
where the factor $\hat W^\dagger(n')\, \hat W(n')$ was inserted in the last step to produce the loop variable. 

\subsubsection*{Case 3: The electric operator \texorpdfstring{$E_L^a(\ell)$}{Ea(l)} is located on a link that is part of the returning sub-path of loop \texorpdfstring{$P(\kappa)$}{P(kappa)}. }
In this case one finds (being much less verbose):
\begin{align}
        & [\hat E_L^a (\ell(n',i')), \hat X(\kappa(n,i))] \\
    & \quad =  -\hat X(\kappa(n, i)) \left( \hat W(n')T^a \hat W^\dagger(n') \right)
    \,.\nonumber
\end{align}

\subsubsection*{Case 4: The electric operator \texorpdfstring{$E_L^a(\ell)$}{Ea(l)} is located on a link that is part of both the outgoing and returning sub-paths. }
This is just the sum of cases 2 and 3
\begin{align}
    & [\hat E_L^a (\ell(n',i')), \hat X(\kappa(n,i))] \\
    & \quad =  \left( \hat W(n')T^a \hat W^\dagger(n') \right) \hat X(\kappa(n,i))\nonumber\\
    & \quad \quad - \hat X(\kappa(n,i)) \left( \hat W(n')T^a \hat W^\dagger(n') \right)
    \,.\nonumber
\end{align}

\subsection{Transported Electric Operators}

The results of the previous section are fairly simple, but can be simplified even further, by noticing that the objects
$W T^a W^\dagger$ appearing in the above commutators are the parallel transport of the adjoint representation object $T^a$ from site $n'$ to the origin along path $\mathcal{P}(n')$. 
One can define parallel-transported electric field operators as\footnote{These operators are defined with respect to the left electric operator, but one could also choose a definition with respect to the right electric operators.}
\begin{align}
    \hat{\mathcal{J}}(\ell(n,i)) &\equiv \hat W^\dagger(n) \hat E_L(\ell(n,i)) \hat W(n)
    \,.
\end{align}
The above equation is expressed in the defining representation -- that is, all objects appearing are $2 \times 2$ matrices (of operators). The content of this equation can equivalently be expressed in the adjoint representation via:
\begin{align}
    \hat{\mathcal{J}}^a(\ell(n,i)) &= 2 \Tr\left[\hat{\mathcal{J}}(\ell(n,i))\, T^a \right]\nonumber\\
    &= 2 \Tr\left[\hat W^\dagger(n) T^b \hat E^b_L(\ell(n,i)) \hat W(n)\, T^a \right]\nonumber\\
    &= 2 \, \hat E_L^b(\ell(n,i))\Tr\left[\hat W^\dagger(n) T^b \hat W(n)\, T^a \right]\nonumber\\
    &= \hat E_L^b(\ell(n,i)) \,\hat W^{1}_{ba}(n)
    \,,
\end{align}
where the operator in the adjoint representation $\hat W^{(1)}_{ab}(n)$ is defined similarly to \cref{eq:adjoing_def} as
\begin{align}
    \hat W^{1}_{ab}(n) \equiv 2 \Tr\left[ \hat W^\dagger(n) T^a \hat W(n) T^b\right]
    \,.
\end{align}
The transported operators $\hat{\mathcal{J}(\ell)}$ have the property that they transform as though located at the origin under gauge transformations. 
For instance
\begin{align}
\label{eq:global_gauge_J}
    \hat{\mathcal{J}}(\ell(n,i)) &\to \Omega(n_0) \hat W(n)^\dagger \cancel{\Omega^\dagger(n) \Omega(n)} \hat E_L(\ell(n,i))\nonumber\\
    & \qquad \times \cancel{\Omega(n)^\dagger\Omega(n)} \hat W(n) \Omega^\dagger(n_0)\nonumber \\
    & \quad = \Omega(n_0) \hat{\mathcal{J}}(\ell(n,i)) \Omega(n_0)^\dagger
    \,.
\end{align}
In addition, they have been chosen to have a simple effect on the link variables. 
One can follow through the same cases as before.
\subsubsection*{Case 1}
As before, the commutator vanishes
\begin{align}
    [\hat{\mathcal{J}}^a(\ell), \hat X(\kappa)] = 0
    \,.
\end{align}

\subsubsection*{Case 2}
One can easily compute the commutator. Beginning with
\begin{align}
    [ &\hat{\mathcal{J}}^a(\ell(n',i')), \, \hat X(\kappa(n,i))] \nonumber\\
    & \quad =  \hat{W}^1_{ba}(n') [ \hat{E}^b_{L}(\ell(n',i')),\, \hat X(\kappa(n,i))]\,,
\end{align}
and using the previously computed commutator of \cref{eq:case2} we have
\begin{align}
    & \quad =  \hat{W}^1_{ba}(n') \hat W(n') T^b \hat W^\dagger(n') \hat X(\kappa(n,i)) \nonumber\\
    & \quad = \cancel{ \hat W(n') }  \cancel{ \hat W^\dagger(n') } T^a \cancel{ \hat W(n') } \cancel{ \hat W^\dagger(n') } \hat X(\kappa(n,i)) \nonumber\\
    & \quad = T^a \hat X(\kappa(n,i)) \,,
\end{align}
where we have used the identity $\hat W^{1}_{ba} T^b = \hat W^\dagger T^a \hat W$, which can easily be derived using the Fierz identity of the generators $T^a$. 

\subsubsection*{Case 3}
Using the same relations as for Case 2, one finds
\begin{align}
    [\mathcal{J}^a(\ell), \hat X(\kappa)] = -\hat X(\kappa)T^a
    \,.
\end{align}

\subsubsection*{Case 4}
The final case is again a combination of cases 2 and 3:
\begin{align}
    [\hat{\mathcal{J}}^a(\ell), \hat X(\kappa)] = T^a \hat X(\kappa)-\hat X(\kappa)\,T^a
    \,.
\end{align}

These formally appear to be the same as the defining commutation relations among the original link operators \cref{eq:EUcommutators}, modulo a minus sign. One is then tempted to conclude that the operators $\hat{\mathcal{J}}_L^a$ and $\hat{\mathcal{J}}_R^a$ can be identified with the left and right generators of the Lie algebras associated with the loop operators $\hat X$. However, because each $\mathcal{J}^a$ acts  on many different loops, this identification is premature.

\subsection{Relationship to the Loop Electric Operators}
Now let $\mathcal{E}^a_L(\kappa)$ and $\mathcal{E}^a_R(\kappa)$ be the left and right generators associated with loop operator $X(\kappa)$, so that they only generate rotations on the specified loop variable and no others
\begin{align}\label{eq:loop_electric_ops}
    [\hat{\mathcal{E}}_L^a(\kappa), \hat X(\kappa')] &= T^a \hat X(\kappa) \delta_{\kappa,\kappa'} \\[10pt]
    [ \hat{\mathcal{E}}_R^a(\kappa), \hat X(\kappa')] &= \hat X(\kappa) T^a \delta_{\kappa,\kappa'}
    \,.
\end{align}
These reproduce the canonical commutations between independent link operators, and are the variables in terms of which we wish to write our Hamiltonian.

Given the commutation relations obtained in the previous section, the operators $\hat{\mathcal{J}}^a(\ell)$ can be related to the loop electric operators defined in \cref{eq:loop_electric_ops}. 
Let $t_{+}(\ell)$ be the set of all physical links $\kappa$ such that $\ell$ is contained in path $P(\kappa)$ as a positive link. 
Similarly let $t_{-}(\ell)$ be the set of all physical links $\kappa$ such that $\ell$ is contained in path $P(\kappa)$ as a negative link. 
Then using the results of the previous section one we identify that
\begin{align}
\label{eq:Ja_relation}
    \hat{\mathcal{J}}^a(\ell) = \sum_{\kappa \in t_+(\kappa)} \hat{\mathcal{E}}^a_{L}(\ell) - \sum_{\kappa \in t_{-}(\kappa)} \hat{\mathcal{E}}^{a}_{R}(\ell)
    \,.
\end{align}
This expression is easy to apply to differing dimensions and loops conventions. 

\subsection{The Hamiltonian}
The magnetic Hamiltonian can easily be obtained using \cref{eq:P_through_X}, and one finds
\begin{align}
\label{eq:HB_maxTree}
    H_B = \frac{1}{2g^2 a}\sum_p \Tr \left( I - \prod_{\kappa\in p} \hat X(\kappa)^{\sigma(\kappa)} \right) + \mathrm{h.c.}
    \,,
\end{align}
which involves at most four loop operators $\hat X(\kappa)$ for each plaquette $p$.

To write the electric Hamiltonian, one starts from its expression in terms of the original link operators as given by \cref{eq:HE}.
Since the adjoint representation of the parallel transporters $\hat W^{1}_{ab}$ are orthogonal $\hat W^{1}_{ab} \hat W^{1}_{ac} = \delta_{bc} \mathds{1}$, one immediately finds
\begin{align}
    \hat E_\ell^2 = 
        \hat{\mathcal{J}}^2_\ell
        \,.
\end{align}
Combining this with \cref{eq:Ja_relation} one finds
\begin{align}
\label{eq:HE_maxTree}
    H_E = \frac{g^2}{2 a} \sum_{\ell} \left( \sum_{\kappa \in t_+(\ell)} \hat{\mathcal{E}}^a_{L\kappa} - \sum_{\kappa \in t_{-}(\ell)} \hat{\mathcal{E}}^{a}_{R\ell} \right)^2
\end{align}
It is not difficult to turn the implicit restricted sums into explicit sums by inspecting the lattice and taking into account the specific conventions of the maximal tree and loop variables.

\bibliographystyle{unsrt}
\bibliography{refs}
\end{document}